\documentclass[nofootinbib,noeprint,aps,prl,reprint,superscriptaddress]{revtex4-1}

\usepackage{tabularx}
\usepackage{lmodern}
\usepackage[T1]{fontenc}
\usepackage{color}
\usepackage{textcomp}
\usepackage{siunitx}
\usepackage{xspace}
\usepackage{mathtools}
\usepackage[dvipsnames]{xcolor,colortbl}
\usepackage{fdsymbol}
\usepackage{amsmath}

\DeclarePairedDelimiter\abs{\lvert}{\rvert}%
\DeclarePairedDelimiter\norm{\lVert}{\rVert}%

\makeatletter
\let\oldabs\abs
\def\abs{\@ifstar{\oldabs}{\oldabs*}}
\let\oldnorm\norm
\def\norm{\@ifstar{\oldnorm}{\oldnorm*}}
\makeatother

\newcommand{\ie}{i.\,e.\xspace}

\newcommand{\sio}{SiO$_2$\xspace}
\newcommand{\is}{I_\text{S}}
\newcommand{\vsd}{V_\text{SD}}
\newcommand{\vbg}{V_\text{BG}}
\newcommand{\ic}{I_\text{SW}}

\newcommand{\ir}{I_\text{R}}

\newcommand{\icrn}{I_\text{C}R_\text{N}}

\newcommand{\dvdi}{\partial{V_\text{SD}}/\partial{I_\text{S}}}

\newcommand{\didvb}{\partial{I_\text{D}}/\partial{V_\text{SD}}}

\newcommand{\bc}{B_\text{C}}
\newcommand{\bcs}{B_{\text{C}\star}}
\newcommand{\bcpe}{B_{\text{C}\perp}}
\newcommand{\bcpa}{B_{\text{C}\parallel}}

\newcommand{\bcal}{B_\text{C,Al}}
\newcommand{\bcsal}{B_{\text{C}\star,\text{Al}}}
\newcommand{\bcpeal}{B_{\text{C}\perp,\text{Al}}}
\newcommand{\bcpaal}{B_{\text{C}\parallel,\text{Al}}}
\newcommand{\bcx}{B_\text{C,X1}}
\newcommand{\bcsx}{B_{\text{C}\star,\text{X1}}}
\newcommand{\bcpex}{B_{\text{C}\perp,\text{X1}}}
\newcommand{\bcpax}{B_{\text{C}\parallel,\text{X1}}}
\newcommand{\bcxx}{B_\text{C,X2}}
\newcommand{\bpe}{B_\perp}
\newcommand{\bpa}{B_\parallel}
\newcommand{\bs}{B_\star}
\newcommand{\bz}{B_\text{Z}}
\newcommand{\tc}{T_\text{C}}
\newcommand{\tcal}{T_\text{C,Al}}
\newcommand{\tcx}{T_\text{C,X1}}
\newcommand{\tcxx}{T_\text{C,X2}}

\newcommand{\ggavg}{\langle G_\text{G}\rangle}

\newcommand{\goavg}{\langle G_\text{O}\rangle}
\begin{document}

\title{Hard superconducting gap and diffusion-induced superconductors in Ge-Si nanowires}



\author{Joost Ridderbos}
\affiliation{MESA+ Institute for Nanotechnology, University of Twente, P.O. Box 217, 7500 AE Enschede, The Netherlands}
\author{Matthias Brauns}
\affiliation{MESA+ Institute for Nanotechnology, University of Twente, P.O. Box 217, 7500 AE Enschede, The Netherlands}

\author{Jie Shen}
\affiliation{QuTech and Kavli Institute of Nanoscience, Delft University of Technology, 2600 GA Delft, The Netherlands}
\author{Folkert K. de Vries}
\affiliation{QuTech and Kavli Institute of Nanoscience, Delft University of Technology, 2600 GA Delft, The Netherlands}

\author{Ang Li}
\affiliation{Department of Applied Physics, Eindhoven University of Technology, Postbox 513, 5600 MB Eindhoven, The Netherlands}
\author{Sebastian K{\"o}lling}
\affiliation{Department of Applied Physics, Eindhoven University of Technology, Postbox 513, 5600 MB Eindhoven, The Netherlands}
\author{Marcel A. Verheijen}
\affiliation{Department of Applied Physics, Eindhoven University of Technology, Postbox 513, 5600 MB Eindhoven, The Netherlands}


\author{Alexander Brinkman}
\affiliation{MESA+ Institute for Nanotechnology, University of Twente, P.O. Box 217, 7500 AE Enschede, The Netherlands}
\author{Wilfred G. van der Wiel}
\affiliation{MESA+ Institute for Nanotechnology, University of Twente, P.O. Box 217, 7500 AE Enschede, The Netherlands}
\author{Erik P. A. M. Bakkers}
\affiliation{Department of Applied Physics, Eindhoven University of Technology, Postbox 513, 5600 MB Eindhoven, The Netherlands}
\author{Floris A. Zwanenburg}
\email[Corresponding author, e-mail: ]{f.a.zwanenburg@utwente.nl}
\affiliation{MESA+ Institute for Nanotechnology, University of Twente, P.O. Box 217, 7500 AE Enschede, The Netherlands}


\date{\today}

\begin{abstract}
We show a hard superconducting gap in a Ge-Si nanowire Josephson transistor up to in-plane magnetic fields of $250$~mT, an important step towards creating and detecting Majorana zero modes in this system. A hard  gap requires a highly homogeneous tunneling heterointerface between the superconducting contacts and the semiconducting nanowire. This is realized by annealing devices at $180$~$^\circ$C during which aluminium inter-diffuses and replaces the germanium in a section of the nanowire. \textcolor{black}{Next to Al, we find a superconductor with lower critical temperature ($\tc=0.9$~K) and a higher critical field ($\bc=0.9-1.2$~T). We can therefore selectively switch either superconductor to the normal state by tuning the temperature and the magnetic field and observe that the additional superconductor induces a proximity supercurrent in the semiconducting part of the nanowire even when the Al is in the normal state.} In another device where the diffusion of Al rendered the nanowire completely metallic, a superconductor with a much higher critical temperature ($\tc=2.9$~K) and critical field ($\bc=3.4$~T) is found. \textcolor{black}{The small size of these diffusion-induced superconductors inside nanowires may be of special interest for applications requiring high magnetic fields in arbitrary direction.}

\end{abstract}




\maketitle

\section{Introduction}
\begin{figure} 
  \includegraphics[width=1.0\linewidth]{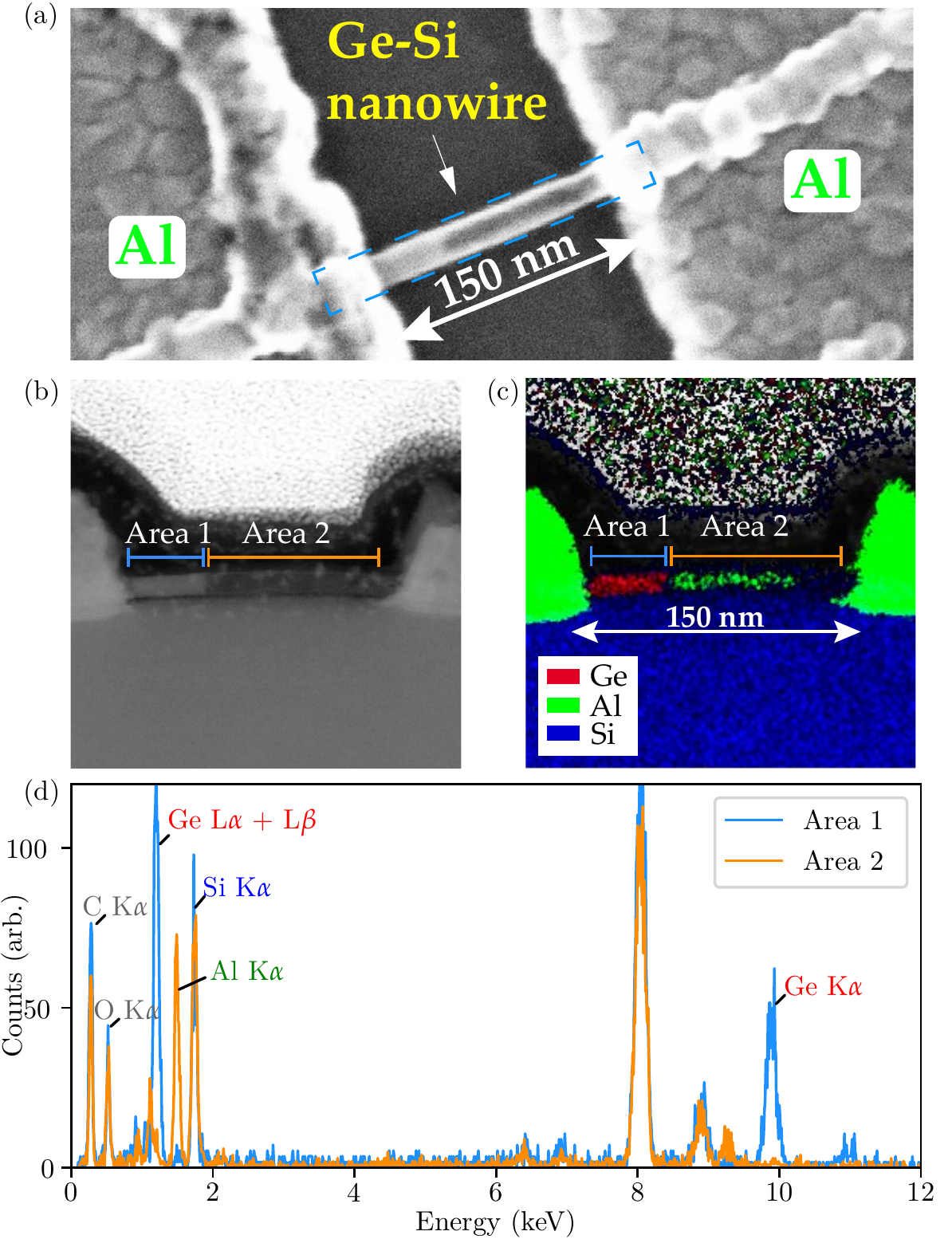}
  \caption{\textbf{Al-Ge inter-diffusion in device A:} a) Top view SEM image of the device showing a Ge-Si nanowire between two Al contacts. In the right part of the nanowire, a slightly darker contrast is observed (see Supplementary Info Fig.~2 for SEM images showing this effect in several devices). The blue dashed line shows the approximate location of the TEM lamella. b) Cross-sectional HAADF-STEM image of the same device. The same contrast difference as in a) is observed. c) HAADF/STEM image with combined EDX data for elements Ge, Al and Si (see Supplementary Info Fig.~S1 for separate images). d) EDX spectrum for Area 1 and Area 2 as defined in c).}
  \label{fig1}
\end{figure}

The discovery that Majorana fermions offer a route towards an inherently topologically protected fault-tolerant quantum computer~\cite{Read1999,DasSarma2005,Nayak2008} marked the beginning of a quickly growing field of research to achieve their experimental realization. Majorana fermions require a topological superconducting material, which in practice can be realized by coupling a conventional $s$-wave superconductor 
to a 1-dimensional nanowire with high spin-orbit coupling and $g$-factor~\cite{Kitaev2000,Oreg2010,Lutchyn2010,Lutchyn2018}. Signatures of Majorana fermions are expected to arise as a conductance peak at zero bias and finite magnetic fields. The first reports showing these zero-bias conductance peaks in InAs and InSb nanowires~\cite{Das2012,Mourik2012b,Deng2012,Lee2012a,Finck2013,Churchill2013,Deng2015a} suffered from sizeable sub-gap conductivity attributed to inhomogeneities in the nanowire-superconductor interface~\cite{Takei2013,Cole2015}. The resulting quasiparticle poisoning decoheres Majorana states since they will participate in braiding operations~\cite{Rainis2012,Higginbotham2015,Albrecht2017}, and additionally obscure the Majorana signatures at zero energy. Strong efforts have been made to improve these interfaces, \ie, induce a hard gap, using epitaxially grown Al~\cite{Chang2015c,Kjaergaard2016a} or specialized surface treatments methods~\cite{Zhang2017,Gul2017}, resulting in a much better resolved Majorana signatures~\cite{Higginbotham2015,Albrecht2016,Deng2016,Gul2018}.
\paragraph{}
In contrast to the group III-V materials used in most previous work, we use Ge-Si core-shell nanowires consisting of a mono-crystalline Ge $\langle110\rangle$ core with a diameter of $\sim$$15$~nm, and a Si shell thickness of $2.5$~nm covered by a native \sio. Coherent strain in the defect-free crystal structure results in high hole-mobilities~\cite{Conesa-Boj2017}. The electronic properties of the one-dimensional hole gas localized in the Ge core~\cite{Lu2005a,Zhang2010} makes them a candidate for observing Majorana fermions~\cite{Maier2014,Thakurathi2017}, and their interaction with a superconductor is still relatively unexplored~\cite{Xiang2006b,Su2016,Ridderbos2017,DeVries2018,Ridderbos2019}. These wires are predicted to have a strong first-order Rashba type spin-orbit coupling~\cite{Kloeffel2011} which, together with the \emph{g}-factor~\cite{Maier2013,Brauns2015}, is tunable by electric fields. Our devices consist of a nanowire channel with superconducting Al source and drain placed on an oxidized Si substrate (for more detailed information about the fabrication process see Supplementary Information SI~1). We focus on two devices where an essential thermal annealing process results in inter-diffusion between Al in the contacts and Ge in the nanowire channel. Device A is an electric-field tunable Josephson junction~\cite{Ridderbos2017,Ridderbos2019} as shown in Fig.~\ref{fig1}a, while in device B the whole semiconducting nanowire channel has been metalized and we suspect Al has largely replaced the semiconductor. 
\paragraph{}
\textcolor{black}{The electric field dependence of Device A has already been extensively studied in Ref.~\cite{Ridderbos2017}, where the main result was the observation of two distinct regimes: a highly transparect regime with a near ideal $\icrn$ product in accumulation, and a tunneling regime with few-hole occupancy where supercurrent only appears at the charge degeneracy points. In this work, we extend on this by investigating the magnetic field dependence of the transport properties in both regimes.} 
\paragraph{}
\paragraph{}
\textcolor{black}{To gain insight into the microscopic properties of the superconductor-semiconductor interfaces, we start by} investigating Device A using high-angle annular dark field (HAADF) - scanning transmission electron microscopy (STEM) in combination with energy-dispersive X-ray spectroscopy (EDX). We find strong indications that the additional superconductor, as well as the highly homogeneous superconductor-nanowire interface arises during the thermal annealing process where Al inter-diffuses with the material in the semiconducting nanowire. In the second part, we map the switching current $\ic$ as a function of critical field $\bc$ and critical temperature $\tc$ of device A and B, which clearly shows an additional superconducting phase in both devices. In the final part we investigate the hardness of the superconducting gap in the semiconducting nanowire of device A, by means of electronic transport measurements near depletion~\cite{Gul2017,Chang2015c} and observe that the conductance in the gap is suppressed by a factor $\sim1000$.

\section{Al-Ge inter-diffusion}

\begin{figure*}
  \includegraphics{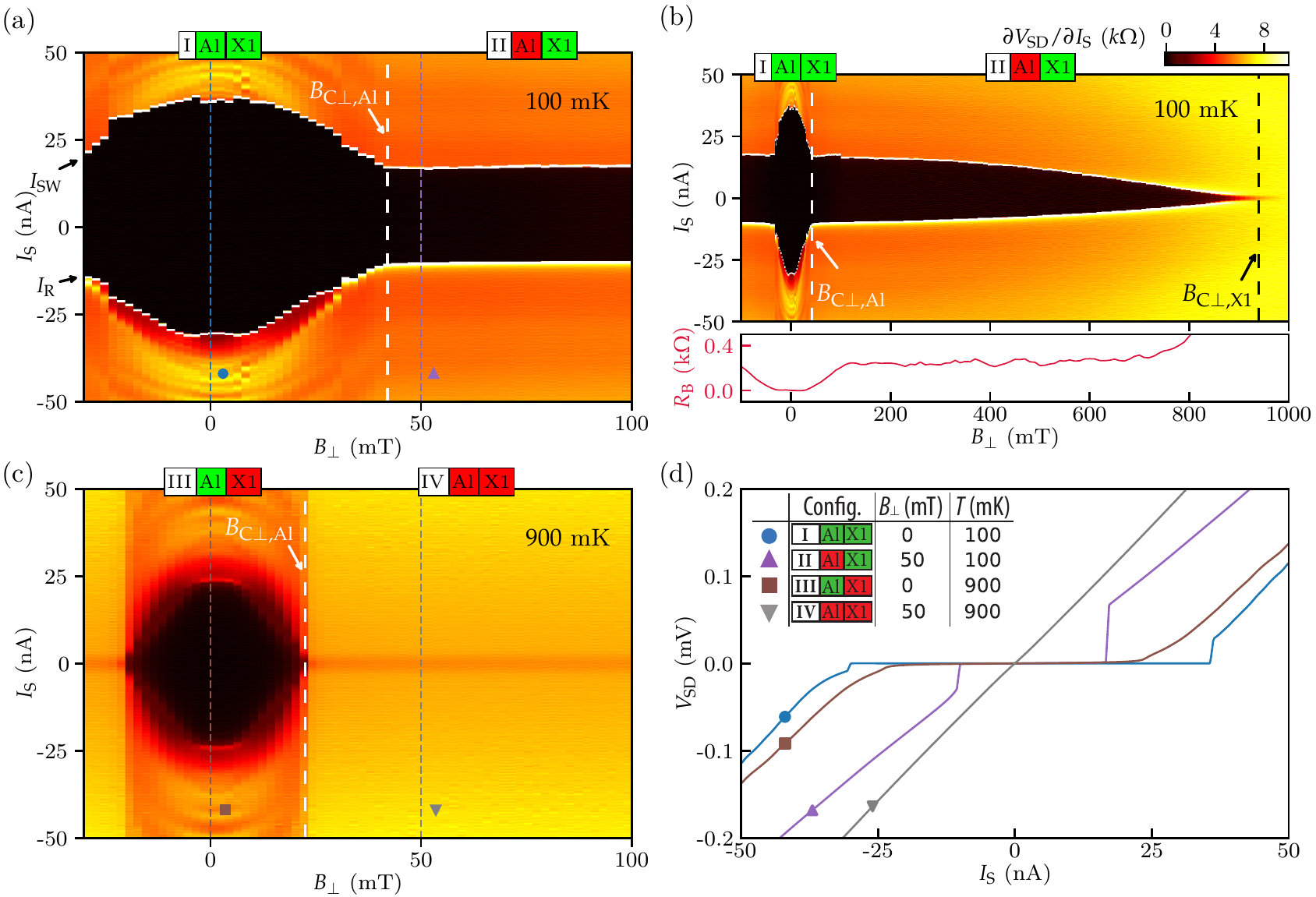}
  \caption{\textbf{Device A: Josephson junction with two superconductors:} a) Differential resistance $\dvdi$ versus $\is$ and $B_\perp$ taken at $T=100$~mK. Black region corresponds to superconductivity. The white dashed line indicates $\bcpeal$. Arrows indicate $\ic$ and $\ir$. b) Top panel: Same as a) for a larger range of $\bpe$. The vertical black dashed line indicates $\bcpex$. Bottom panel: Horizontal cross-section showing $\dvdi$ vs $\bpe$ taken at $\is=0$. The color scale also applies to a) and b). c) Same as a) taken at $T=900$~mK. d) Combinations of Al and X$1$ in the superconducting (green boxes) / normal (red boxes) state are numbered as configurations I-IV. Linecuts showing $\vsd$ versus $\is$ taken for each configuration at the corresponding symbols in (a) and (c). Inset: table summarizing the configurations and values of $\bpe$ and $T$ for the respective linecuts. In all figures $\vbg=-4.7$~V and $\is$ is swept from negative to positive bias.}
  \label{fig2}
\end{figure*}

\textcolor{black}{To investigate the effects of the annealing on the stoichiometric composition of the nanowire channel, a TEM lamella was made along the nanowire axes of device A as indicated in Figure~\ref{fig1}a. \textcolor{black}{We first apply a stack of protective \sio and Pt layers and subsequently create the TEM lamella using a standard focused ion beam lift-out protocol.} This allows us to perform an analysis on the cross-section of the device as can be seen in Fig.~\ref{fig1}b. In both Fig.~\ref{fig1}a and b, a smaller region (Area 1) with higher contrast on the left, and a bigger region with lower contrast on the right (Area 2) can be observed. Fig.~\ref{fig1}c shows the resulting EDX signals in these regions for the elements Ge, Si and Al and we observe a clear distinction: in Area 1 we observe a strong Ge signal while in Area 2 the signal is dominated by Al.
\paragraph{}
In Figure~\ref{fig1}d we show the integrated EDX spectra for both areas. When comparing the two areas, we observe that in Area 2 the Ge~$L\alpha$, Ge~$L\beta$ and Ge~$K\alpha$ signals fall below the detection limit. \textcolor{black}{As is the convention in EDX analysis, $L$ and $K$ denote the orbital to which an electron decays in a picture where $K$, $L$, and $M$ are the outer atomic orbitals, while $\alpha$ and $\beta$ indicate whether it decays from the first or second higher orbital.}
The Al~$K\alpha$ signal shows the opposite behavior, implying that Ge has been replaced by Al in Area 2. The counts for elements O, C and Si remain equal in both areas (see also Supplementary Information Fig.~S1). As we will discuss in the following section, the superconductor in Area 2 has profoundly different properties from the Al contacts and we therefore refer to it as X$1$. \textcolor{black}{Inter-diffusion has also taken place below the left contact without reaching the channel, although this is not evident from the TEM data. Instead, we conclude this from transport data in the next section (Fig.~\ref{fig2} and Supplementary Information Fig.~S3).} As a side-note, we cannot observe the effects of the inter-diffusion process on the Si shell, since the Si signal is dominated by the \sio that covers the substrate.}
\paragraph{}
\textcolor{black}{An in-depth study on the thermally induced inter-diffusion process between Al and pure Ge $\langle111\rangle$ nanowires, a highly similar system to ours, has been performed in Refs.~\citenum{Kral2015,ElHajraoui2019}. Here, in-situ monitoring of the metal front inside the nanowires at various temperatures reveals that the velocity of propagation as a function of the length of the metalized nanowire segment, is volume-diffusion limited, and possibly surface-diffusion limited, with the Al forming a mono-crystalline face-centered-cubic crystal inside the nanowire. The metal front forms an atomically sharp interface and no inter-metallic phase is found in the metalized nanowire segment, i.e., the Ge is transported out of the wire into the Al contacts. These observations are explained by a 15 orders of magnitude lower diffusion constant for Al in Ge than for Ge in Al~\cite{Gale2003,Villars2012}. Furthermore, the initial start of the diffusion reaction is governed by the respective activation energies ($121.3$~kJ/mol for Ge in Al, $332.8$~kJ/mol Al in Ge~\cite{Gale2003,Villars2012}) and may depend on the specific atomic arrangement of the initial nanowire-Al interface, explaining the variation in the starting time of the diffusion reaction, even for two separate contacts on the same wire. These findings largely correspond to our observations on Ge-Si core-shell nanowires and gives an explanation for the asymmetry in our contacts (see Supplementary Figure~S2 for SEM images of partly and fully metalized nanowires) , as well as the variation in device properties.} 

\section{Two superconductors in a nanowire Josephson junction}
\begin{figure*}
  \includegraphics{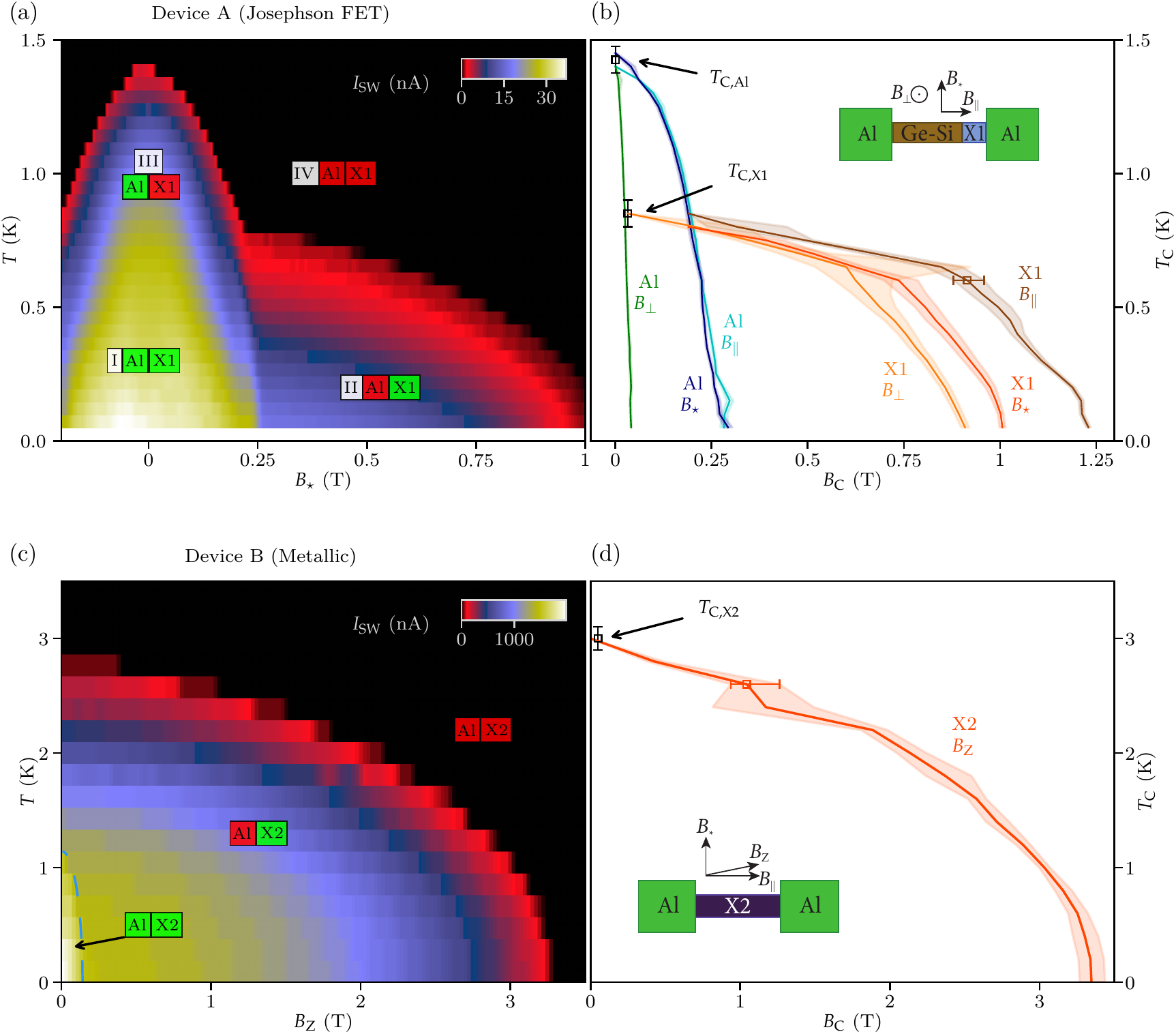}
  \caption{\textbf{$\ic$, $\tc$ and $\bc$ of a Josephson FET (device A) and a metallized nanowire device (device B):} a) $\ic$ versus $T$ and $\bs$ for the Josephson FET (device A). The green (red) boxes indicate whether the material is superconducting (normal) and show the configurations~\textbf{I-IV} as defined in the main text. b) $\tc$ vs $\bc$ for Al and X$1$ for three main field axis $\bpe$, $\bpa$ and $\bs$ as illustrated by the inset. Curves are extracted from plots such as a) (see main text). c) $\ic$ versus $T$ and $\bz$ for the completely metallized nanowire (device B) consisting of alloy X$2$. The green (red) boxes indicate three possible configurations. For the configuration where Al is superconducting (for $\bz<300$~mT and $T<1$~K) an enhancement of $\ic$ can be observed as denoted by the blue dotted line. d) $\tc$ vs $\bc$ for X$2$ extracted from c). Inset shows the in-plane $\bz$ field direction which is rotated \textasciitilde$\ang{10}$ with respect to the nanowire. $\bz$ corresponds to the z-axis of the vector magnet, the only axis capable of fields $>1$~T. In both b) and d), the vertical error bar represents an uncertainty in $\tc$ of $2$~\% and shaded areas are standard deviations in $\bc$ from fits.}
  \label{fig3}
\end{figure*}

In Fig.~\ref{fig2}a we show a magneto-spectroscopy of device A, the Josephson junction: we plot the differential resistance $\dvdi$ versus the sourced current $\is$ and the out-of-plane magnetic field $\bpe$ (see illustration in Fig.~\ref{fig3}b) while sweeping $\is$ from negative to positive current. The backgate $\vbg$ is fixed at $-4.7$~V where multiple subbands contribute to transport and the junction is highly transparent~\cite{Ridderbos2017}. The superconducting region (black) is bounded by  $\ir<\is<\ic$ with $\ir$ the retrapping current at negative bias and $\ic$ the switching current at positive bias. Upon increasing $\bpe$ from $0$, $\ic$ decreases gradually until aluminum becomes normal at the critical out-of-plane field $\bcpeal\approx40$~mT after which a finite $\ic$ remains. For all $\bpe$, $\ic>\ir$ \textcolor{black}{indicating that our junction is hysteretic for this particular value of $\vbg$ due to the junction being underdamped~\cite{Ridderbos2017} while additional heating-induced hysteresis can not be excluded~\cite{Courtois2008}}  (see Supplementary Information Fig.~S3a for a gate-dependence of $\ic$ and $\ir$).
\paragraph{}
When increasing $\bpe$ further in Fig.~\ref{fig3}b, $\ic$ slowly decreases and finally disappears. The proximity-induced supercurrent above $\abs{\bcpeal}$ implies the presence of a second superconducting material, X$1$, in or near the nanowire channel with a critical field $\bcpex\approx950$~mT. To confirm that our Al contacts are normal for $\bpe>\abs{\bcpeal}$, we consider the background resistance $R_\text{B}$ in the superconducting region as a function of $\bpe$ in the bottom panel of Fig.~\ref{fig2}b. $R_\text{B}=0$ for $\bpe<\abs{\bcpeal}$, while for $\bpe>\abs{\bcpeal}$ the background resistance gradually increases to $R_\text{B}\approx0.25$~k$\Omega$ attributed to a normal series resistance of the Al contacts. Additionally, the out-of-plane critical field of a separately measured Al lead matches $\bcpeal$ (see Supplementary Information Fig.~S4).
\paragraph{}
In Fig~\ref{fig2}c we show a magneto-spectroscopy at $900$~mK and observe that X$1$ is quenched for all $\bpe$, while Al still induces a supercurrent for $\bpe<\abs{25}$~mT. This shows that X$1$ has a lower $\tc$ and a higher $\bc$ than the Al contacts. Because X$1$ has a higher $\bc$ and a lower $\tc$ than Al, we can selectively switch either superconductor to the normal state, resulting in four possible device configurations~\textbf{I-IV} \textcolor{black}{as illustrated in Fig.~\ref{fig2} and summarized in the inset in Fig.~\ref{fig2}d (a precise set of conditions for each configuration can be found in Supplementary Information Table~S1).} Fig.~\ref{fig2}d shows plots of $\vsd$ versus $\is$ in all four configurations, clearly showing a supercurrent in configuration~\textbf{II} where Al is normal and only X$1$ is superconducting. \textcolor{black}{Since we observe a gate-tunable Josephson current even in configuration~\textbf{II}, we conclude X$1$ is present on both sides of the Ge-Si segment (see Supplementary Information Fig.~S3 for differential resistance maps versus backgate in all four configurations).}

\section{Junction $\ic$ versus $B$ and $T$}
For the observed superconductors and their specific geometries, the critical field and critical temperature are inter-dependent variables and may have a non-trivial relation, the boundaries of the configurations \textbf{I-IV} in terms of $\bc$ and $\tc$ cannot directly be deduced from the data in Fig.~\ref{fig2}. We therefore collect $\ic$ versus $B$ from magneto-spectroscopies for a large number of temperatures and the three main magnetic field axes $\bs$, $\bpe$ and $\bpa$ which are illustrated by the inset in Fig.~\ref{fig3}b. For the in-plane field perpendicular to the nanowire, $\ic$ has two clearly distinct overlapping shapes as a function of $T$ and $\bs$ in Fig.~\ref{fig3}a: The `peak' extending to $T\approx1400$~mK at $B=0$ with a width of $\abs{\bs}\approx250$~mT at $T=50$~mK is attributed to the superconducting state of Al, while the second shape (the `tail'), extending up to $\sim1000$~mT at $T=50$~mK, corresponds to the superconducting phase of X$1$. We can thus map the four configurations in the color plot on the $T$ vs $\bs$ axes.
\paragraph{}
We now extract both the $\tc$-$\bcsal$ and $\tc$-$\bcsx$ curves from Fig.~\ref{fig3}a (see Supplementary Information section SII), \ie, the critical temperature - critical field relation for Al and X$1$, and plot them in Fig.~\ref{fig3}b. We perform the same procedure for field directions $\bpe$ and $\bpa$ (see Supplementary Information Fig.~S5 for $\ic$ versus $T$ and $\bpa$ and $\bpe$).

\begin{table}[htbp]
	\caption{\label{tab:scparam} Maximum values for $\tc$, $\bc$ of Al, X$1$ and X$2$ as determined in Fig.~\ref{fig3}. We take $\tcal(\bc=0)$, $\tcx(\bpe=50~\mathrm{mT})$ and $\bc(T\approx0)$ to obtain their respective maximum values. The BCS superconducting gap is determined as $\Delta=1.764k_\text{B}\tc$~\cite{Tinkham}.}
    \footnotesize
    \centering
    \setlength\tabcolsep{2pt}
    \begin{tabularx}{\columnwidth}{lrrrrr}
		 				& $\tc$ (K) 	& $\Delta$ ($\mu$V) 		& $\bcs$ (mT) 		& $\bcpe$ (mT) 		& $\bcpa$ (mT) 	\\ \hline
    	\textbf{Al} 	& $1.4\! \pm\! 0.05$	& $212\! \pm\! 6$					& $293\! \pm10\! $    	& $41\! \pm\! 2$      	& $282\! \pm\! 10$    \\
    	\textbf{X1} 	& $0.9\! \pm\! 0.05$  & $133\! \pm\! 8$					& $1230\! \pm10\! $  		& $909\! \pm\! 11$    	& $1010\! \pm\! 20$ 	\\ \\ \hline \hline
    	  				& $\tc$ (K)		& {$\Delta$} ($\mu$V) 		& \multicolumn{3}{l}{$B_\text{C,Z}$ (T)} 	\\ \hline
    	\textbf{X2} 	& $2.9\! \pm\! 0.1$   & $441\! \pm\! 14$				& \multicolumn{3}{l}{$3.4\! \pm\! 0.1$} 			\\
    \end{tabularx}
\end{table}

\paragraph{}
In Table.~\ref{tab:scparam} we summarize the maximum $\tc$, the resulting superconducting gap $\Delta$ and $\bc$ in the three field directions for Al and X$1$. Comparing $\bcpeal=41$~mT with $\bcsal=293$~mT and $\bcpaal=282$~mT we notice a factor \textasciitilde{}$7$ difference. This strong anisotropy for the out-of-plane field direction is clearly present in the $\tcal$-$\bcal$ curves in Fig.~\ref{fig3}b and is expected for the large aspect ratio of the 50~nm thick Al contacts. 
\paragraph{}
The $\tcx$-$\bcx$ curves show a less prominent magnetic field anisotropy from which we can roughly deduce the shape of X$1$ by assuming that the normal surface of the material is inversely proportional to the critical field, \ie, a larger superconducting normal-surface requires expelling more flux~\cite{Tinkham}. Using the respective ratios of $\bcsx$, $\bcpex$ and $\bcpax$ we observe that X$1$ is slightly elongated along the nanowire axis, reaffirming the hypothesis that X1 resides in the nanowire channel.
\paragraph{}
We now switch to the completely metalized device B where we believe Al has diffused completely through the channel, effectively making the nanowire a metallic superconductor. Fig.~\ref{fig3}c shows $\ic$ vs $T$ and $\bz$ where the corresponding $\tcxx$-$\bcxx$ relation in Fig.~\ref{fig3}d is obtained by the previously mentioned polynomial fitting method. We see a critical temperature $\tcxx=2.9$~K at $B=0$ and critical field $\bcxx=3.4$~T at $T=50$~mK, both much higher than for X$1$ and the Al contacts. The switching current $\ic=1.5$~$\mu$A is two orders of magnitude higher compared to device A. 
\paragraph{}
When comparing $\tcxx=2.9$~K and $\bcxx=3.4$~T with thin Al aluminium films~\cite{Meservey1971}, we observe X$2$ has equivalent properties of a $\sim3$~nm thick film (in parallel field) and we could conclude that X$2$ is simply a very small cylinder of aluminium inside the nanowire channel. However, for X$1$ with $\tcx=0.9$~K and $\bcx\approx1$~T an equivalent film thickness cannot be defined. \textcolor{black}{Even though no inter-metallic phases were found for annealed pure Ge nanowires in Refs.~\citenum{Kral2015,ElHajraoui2019}, a possible origin of X$1$ is the formation of a Al-Si/Ge alloy in our core-shell nanowires, albeit with a ratio of semiconductor to Al below that of our EDX detection limit.} In literature, certain stoichiometric compositions indeed result in a lower $\tc$ than for pure Al~\cite{Deutscher1979,Lesueur1988} and in fact, one can get alloys with a $\tc$ ranging from $0.5$~K up to $11$~K by various methods~\cite{Deutscher1971,Tsuei1974,Kuan1982,Chevrier1987,Xi1987}. The exact composition of both X$1$ and X$2$ in our Ge-Si core-shell system therefore remains partly speculative \textcolor{black}{and would require a more in-depth study like Ref.~\citenum{ElHajraoui2019}.}
\paragraph{}
To sum up, we observe X$1$ with $\tcx=0.9$~K in a Josephson junction and X$2$ with $\tcxx=2.9$~K in a metallic device, showing that diffusion of Al into Ge-Si nanowires can gives rise to different superconductors with a $\tc$ lower and much higher than that of the Al contacts, both appear as a second superconductor in transport measurements.

\begin{figure*}
  \includegraphics[scale=1]{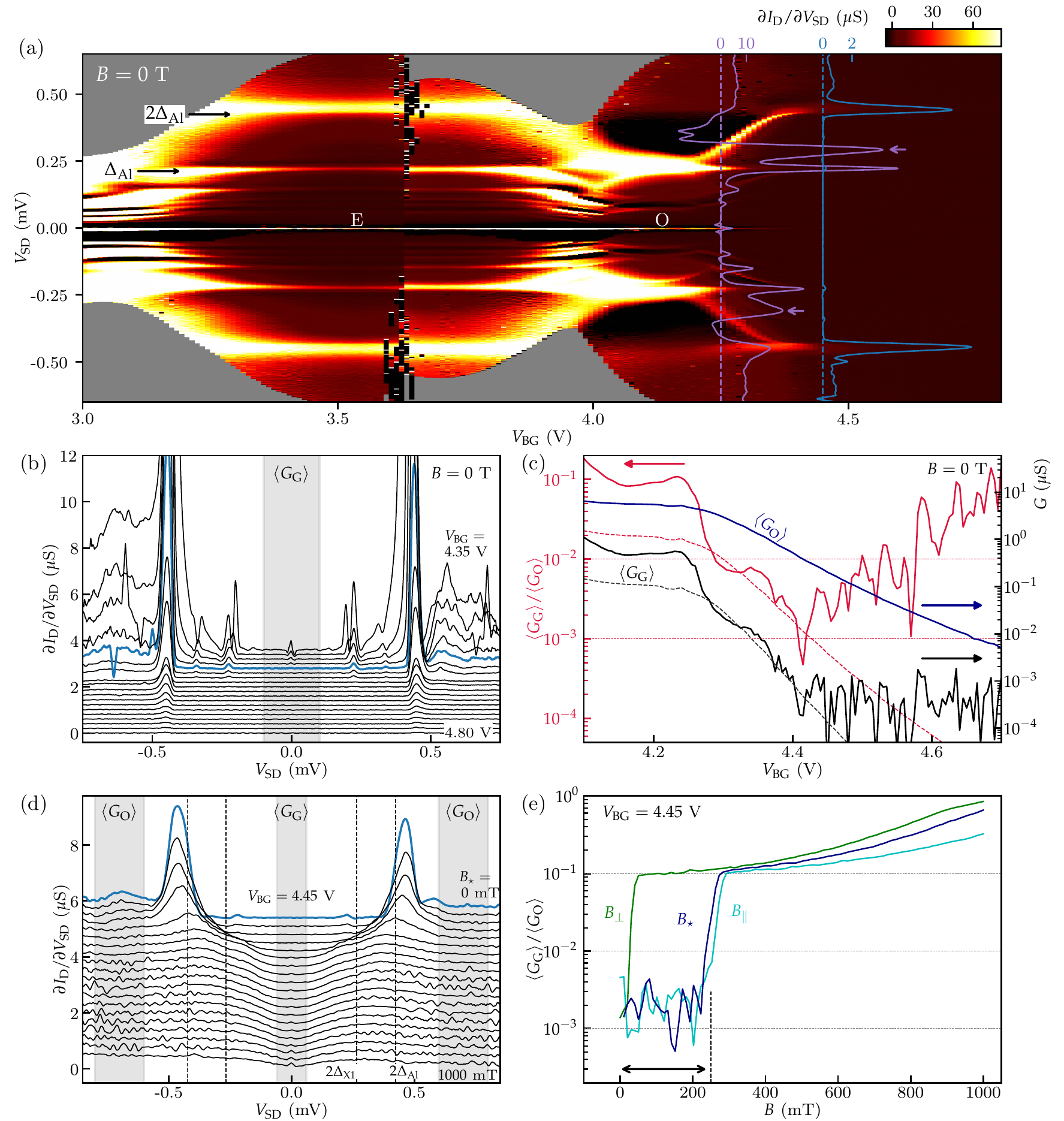}
  \caption{\textbf{Hard superconducting gap in a Ge-Si nanowire Josephson FET (Device A):} a) Differential conductance $\didvb$ vs $\vsd$ and $\vbg$. Odd (O) and even (E) hole occupation are denoted. The first two MAR orders are indicated at $\vsd=2\Delta_\mathrm{Al}$ and $\Delta_\mathrm{Al}$. b) Vertical linecuts from a) showing $\didvb$ vs $\vsd$ at $50$~mV intervals in $\vbg$, curves are offset by 0.2~$\mu$S. c) \textcolor{black}{Averaged in-gap conductance $\ggavg$ (black) and outside-gap conductance $\goavg$ (blue) and the ratio $\ggavg/\goavg$ (red) versus $\vbg$. Dashed curves show theoretical minimal values and  are the result of plotting Eq.~(\ref{eqhg}). For every $\vbg$, $\ggavg$ and $\goavg$ are averaged over a range of $\vsd$ as indicated by the grey area in (b) and the grey dashed lines in Fig.~SI-7 respectively.} d) $\didvb$ vs $\vsd$ for $\bs$ from $0$ to $1000$~mT at $50$~mT intervals. Curves are offset by 0.3~$\mu$S. Dashed lines show the expected position of the quasiparticle peak for $2\Delta_\mathrm{Al}$ ($2\Delta_\mathrm{X1}$) at $B=0$. e) Ratio $\ggavg/\goavg$ for the three main field axes $\bpe$, $\bpa$ and $\bs$ at $\vbg=4.45$~V (blue line in (a), (b) and (c)). Ranges in $\vsd$ where $\ggavg$ and $\goavg$ are extracted are shown as grey areas in d).}
  \label{fig4}
\end{figure*}

\textcolor{black}{\section{Tunneling regime of the Josephson FET}}
We now focus on device A and tune $\vbg$ to a regime where the nanowire is near depletion. Fig.~\ref{fig4}a shows the differential conductance $\didvb$ versus the source-drain voltage $\vsd$ and the backgate voltage $\vbg$. We notice a zero-bias conductance peak as the result of a finite Josephson current and a prominent multiple Andreev reflection (MAR) pattern showing as horizontal lines of increased conductance for $\vbg=3$ to $4$~V. The reduced barrier transparency near depletion confines charges in the nanowire channel, and allows us to see odd and even charge occupation in a quantum dot in the wire~\cite{Ridderbos2017} supported by a Kondo peak on the odd transitions~\cite{Ridderbos2017,Kim2013} (see Supplementary Information, Fig.~S6). Above $\vbg=4.4$~V, the MAR and zero-bias peak disappear, while the onset of quasiparticle transport is visible at the superconducting gap at $\vsd=\pm{2\Delta_\text{Al}}$. This trend is also present in the $\didvb$ linecuts for $\vbg$ between $4.35$ and $4.80$~V in Fig.~\ref{fig4}b.
\paragraph{}
\textcolor{black}{In Fig.~\ref{fig4}a between $\vbg=4.2$~V and $\vbg=4.4$~V we observe a conductance peak in both bias directions smoothly moving from $\abs{\vsd}=\Delta_\text{AL}$ to $\abs{\vsd}=2\Delta_\text{AL}$ when going from the odd to the even occupancy, which we attribute to an Andreev bound state (ABS). Additional evidence for an ABS presents itself in the form of a region of negative differential conductance in the odd occupancy between $\vsd=\Delta_\text{AL}$ and $\vsd=2\Delta_\text{AL}$~\cite{Pillet2010,Gramich2017a}, as highlighted by the purple linecut at $\vbg=4.25$~V in Fig.~\ref{fig3}a. Tunnel spectroscopy on an ABS requires asymmetric opaque tunnel barriers where the most opaque barrier probes the ABS~\cite{Kim2013}. A barrier asymmetry in our devices can indeed be expected, since the final interface properties are determined by microscopic details on the Al-nanowire interface during annealing. For lower $\vbg$ our barriers quickly become highly transparent~\cite{Ridderbos2017} and we therefore only observe the ABS signature near depletion.
\paragraph{}
In contrast to the bias-symmetric MAR features, the asymmetric barriers show up in the intensity of the ABS signatures (see the arrows on the purple linecut in Fig.~\ref{fig4}a). Depending on the bias direction, there are two different rate-determining tunnel sequences: (1) tunneling through an opaque barrier onto a single ABS or (2) tunneling from an ABS through an opaque barrier into the Fermi sea. Sequence (2) has a much higher tunnel probability than (1), which results in the observed asymmetry in conductance.} 

\section{Hard superconducting gap}
A measure for the amount of quasiparticle states inside the gap, is the in-gap suppression of conductance also termed as the hardness of the gap. We therefore investigate the ratio $\ggavg/\goavg$ where $\ggavg$ ($\goavg$) is a conductance value inside (outside) the gap averaged over a range of $\vsd$ as shown in Fig.~\ref{fig4}b. \textcolor{black}{$\goavg$ is determined from a similar measurement at higher bias (see Fig.~SI-7), sufficiently far away from $2\Delta_\mathrm{Al}$.} Fig.~\ref{fig4}c shows $\ggavg$, $\goavg$ and the ratio $\ggavg/\goavg$ versus $\vbg$ and we find the conductance is suppressed by a factor $\sim$$1000$ for $\vbg\approx4.4$~V, an order of magnitude higher than previously reported in this system \textcolor{black}{in the same superconductor-normal-superconductor (SNS) configuration~\cite{DeVries2018}.}  
\textcolor{black}{
\paragraph{}
A SNS junction can naively be viewed as two superconductor-normal (SN) junctions in series and the theoretical dependence of $G_\mathrm{G}$ on $G_\mathrm{O}$ can therefore be approximated as~\cite{Beenakker1992a}:
\begin{equation}
G_\mathrm{G,SNS}\approx \frac{G_\mathrm{G,SN}}{2}=\frac{2e^2}{h}\frac{G_\mathrm{O}^2}{\left(\frac{4e^2}{h}-G_\mathrm{O}\right)^2},
\label{eqhg}
\end{equation}
}
\textcolor{black}{and it follows that the equivalent conductance suppression of a SN device is a factor two lower than for a SNS device. We use the averaged $\goavg$ as $G_\mathrm{O}$ and obtain the theoretical minimal in-gap conductance $G_\mathrm{G,SNS}$, as well as the corresponding ratio $G_\mathrm{G,SNS}/G_\mathrm{O}$, shown as dashed lines in Fig.~\ref{fig4}c. We find that above $\vbg=4.25$~V, the measured $\ggavg$ and $\ggavg/\goavg$ closely follow the theoretical curves until the noise limit of our equipment is reached for $\vbg>4.4$~V. This suggests that $\ggavg$ is not dominated by quasiparticle poisoning and that our superconductor-semiconductor interfaces do not facilitate inelastic scattering and have low disorder~\cite{Takei2013}. We note that for these values of $\vbg$, the Ge-Si island is not fully depleted ($\goavg$ still decreases as a function of $\vbg$ and can be fully suppressed) and transport takes place through a tunnel-broadened quantum dot level (see also Ref.~\citenum{Ridderbos2017}). However, the obtained theoretical minimal in-gap conductance should be considered an approximation since we do not take into account any difference in interface transparency between the two contacts.
\paragraph{}
When measured in a SNS configuration, the ratio $\ggavg/\goavg$ gives an upper limit and could in reality be lower}
since it can be increased due to several other reasons than quasiparticle poisoning: (1) for higher $\vbg$, $\ggavg$ is limited by the noise floor of our measurement setup and does not further decrease. The decrease of $\goavg$ now lowers the observed current suppression $\ggavg/\goavg$. (2) For lower $\vbg$, MAR and the zero-bias peak, both characteristic for Josephson junctions, appear as conductance peaks inside the gap which leads to a decreased $\ggavg/\goavg$ \textcolor{black}{(3) The quantum dot in the junction may lead to Fabry-Perot resonances and Kondo-enhanced tunnelling around zero bias (see Fig.~SI-6). SN devices will not exhibit these effects} and may therefore result in a lower ratio $\ggavg/\goavg$ and give a better approximation of the quasiparticle density in the gap. \textcolor{black}{Because of this, we cannot directly compare the current suppression in our device with other work probing the superconducting gap using a single superconducting contact. Nevertheless, the fact that our $\ggavg/\goavg$ is limited by the noise floor our measurement setup suggests that our semiconductor-nanowire interface homogeneity could be comparable to InAs nanowire devices using epitaxial growth techniques~\cite{Chang2015c} or specialized surface treatments~\cite{Gul2017}.}
\paragraph{}
We will now look at the magnetic field dependence of the hardness of the gap. We fix $\vbg$ at $4.45$~V and plot $\didvb$ versus $\vsd$ for several $\bs$ in Fig.~\ref{fig4}d. For increasing $\bs$, the sharp quasiparticle peak at $\vsd=2\Delta_\text{Al}$ reduces in height and broadens up to $\bcsal\approx300$~mT. Above $\bcsal$, we enter configuration~\textbf{II} where only X$1$ is superconducting but which fails to produce a clear second quasiparticle peak at $\sim2\Delta_\text{X1}$. Instead, we see a `soft gap' signature~\cite{Takei2013} persisting up to $\bcsx$ \textcolor{black}{which we attribute to X$1$ having an ill-defined gap due to possible diffusion-induced spatial variations in its stoichiometry or geometry.}
\paragraph{}
In Fig~\ref{fig4}e we plot the ratio $\ggavg/\goavg$ for the three main field directions. The initial ratio is \textasciitilde~$1\cdot10^{-3}$ in configuration~\textbf{I} as defined in Fig.~\ref{fig2} and the gap remains hard until we approach the critical field of Al for the respective field direction as summarized in Table~\ref{tab:scparam} (See Supplementary Information Fig.~S8 for the corresponding differential conductance maps for all three main field axes). The highest field where the gap remains hard, $\bpa\approx250$~mT, is slightly lower than $\bcpaal$ because of the strongly reduced $\Delta_\text{Al}$ at this field. The much softer gap in configuration~\textbf{II} induced by X$1$ leads to a $\ggavg/\goavg\approx1\cdot10^{-1}$ which gradually increases to $1$ approaching $\bcx$. 

\textcolor{black}{Another example of the change in transport properties when Al becomes normal is seen in Fig.~\ref{fig2}a and Fig.~\ref{fig2}c. Here, the fringes in the normal state attributed to multiple Andreev reflections (MAR) are only visible for $\bpe<\bcpeal$. For $\bpe>\bcpeal$, the absence of MAR suggests an increase of inelastic processes due to an ill-defined induced gap or a greatly increased quasiparticle poisoning rate.}
\paragraph{}
The results in Fig.~\ref{fig4}e show that the Al contacts needs to be superconducting in order to observe a hard gap. On the other hand, when only Al is superconducting, \ie, going from configuration \textbf{I} to \textbf{III} we observed no change in $G_\text{G}$ that can be attributed to X$1$ becoming normal (see Supplementary Information, Fig.~S9 for the temperature dependence of the differential conductance at $\vbg=4.45$~V and $B=0$). This suggests that X$1$ does not need to be a superconductor to observe a hard gap as long as the Al contacts proximise the entire junction. \textcolor{black}{This is likely to happen, since the transparency between Al and X$1$ is high, and $\Delta_\text{Al}>\Delta_{\text{X}1}$ indicating a coherence length for X$1$ comparable or larger than for Al, \ie, in the order of $\mu$m~\cite{Gross2005}.} 
\paragraph{}
\textcolor{black}{Previously in this system a soft gap signature using NbTiN contacts has been shown~\cite{Su2016}, as well as hard gap using Al contacts~\cite{DeVries2018}. This work adds an investigation of the superconductor-semiconductor interfaces and their microscopic properties.} 
We therefore revisit Fig.~\ref{fig1}b and c and take a closer look at the interface between the X$1$ and the Ge-Si island. Even though our TEM and EDX resolution prohibits a conclusive statements about the interface properties on an atomic scale, the abrupt change in contrast suggests an upper limit for the interface width of a few nanometer. 
\textcolor{black}{As explained, this observation is supported by Refs.~\citenum{Kral2015,ElHajraoui2019} showing an atomically sharp interface between the Ge and Al segment where both remain crystalline~\cite{Kral2015,ElHajraoui2019}. This type of interface would fit our observation of a hard gap, requiring a defect-free highly homogeneous heterointerface~\cite{Takei2013} \textcolor{black}{and low junction transparency close to depletion. This indicates that the inter-diffusion reaction between Ge and Al} is essential for the observed hard superconducting gap~\cite{Kral2015,ElHajraoui2019}.}
\paragraph{}
\textcolor{black}{Utilizing these interfaces in devices suitable for measuring Majorana fermions in this system~\cite{Maier2014a} would require a high level of control over the inter-diffusion process, i.e., lateral diffusion and metalization of nanowire segments should be prevented. One route would be to perform device annealing while in-situ monitoring of the diffusion process as in Ref.~\citenum{ElHajraoui2019}, or possibly a higher level of control could be achieved by optimizing the annealing process. In addition, one would require thinner Al leads in order to withstand the required in-plane magnetic fields ($>1$~T) to reach the topological phase transition~\cite{Maier2013,Maier2014}.} 
\paragraph{}
\textcolor{black}{With a controlled inter-diffusion reaction, the superconductors X$1$ and X$2$ themselves would also pose as interesting materials since their high $\bc$ in relation to their superconducting gaps, might allow the creation of Majorana fermions in materials where low \emph{g}-factors could be limiting~\cite{Stanescu2013a}. However, more research is required to understand the soft gap induced by X$1$ and to fully explore the possible superconductors, their composition, and formation process.}

\section{Conclusion}
\textcolor{black}{
We have shown that Ge-Si nanowire devices with Al contacts contain additional superconductors after annealing, caused by diffusion of Al into the nanowire channel. We identify two superconductors in two different devices: X$1$ is present in a Josephson FET and X$2$ resides in a metallic nanowire channel. Both X$1$ and X$2$ remain superconducting for magnetic fields much higher than the Al contacts which could be of potential interest for applications where proximity-induced superconductivity is required in high magnetic fields.
\paragraph{}
Close to depletion, the Josephson FET exhibits a hard superconducting gap where the in-gap conductance is suppressed by a factor \textasciitilde$1000$ \textcolor{black}{in a SNS configuration where the in-gap conductance is close to the approximate theoretical minimum. The gap remains hard} up to magnetic fields of $\sim250$~mT. For higher fields, a soft gap remains up to the critical field of X$1$. We can selectively switch Al or X$1$ from the normal to the superconducting state and, together with the TEM and EDX analysis, \textcolor{black}{we believe that the diffusion-induced homogeneous heterointerface between the Ge core and the metalized nanowire segment is key} in obtaining this hard gap. The next challenge is to more precisely control the diffusion of Al which would grant a highly promising system for observing Majorana zero modes~\cite{Maier2014}.}
\section{Acknowledgements}
\begin{acknowledgments}
The authors acknowledge financial support from the Netherlands Organization for Scientific Research (NWO). E.P.A.M.B. acknowledges financial support through the EC Seventh Framework Programme (FP7-ICT) initiative under Project SiSpin No. 323841. Solliance and the Dutch province of Noord-Brabant are acknowledged for funding the TEM facility. This project has received funding from the European Union's Horizon 2020 research and innovation programme under Grant Agreement \#862046.

\end{acknowledgments}


\begin{thebibliography}{60}%
\makeatletter
\providecommand \@ifxundefined [1]{%
 \@ifx{#1\undefined}
}%
\providecommand \@ifnum [1]{%
 \ifnum #1\expandafter \@firstoftwo
 \else \expandafter \@secondoftwo
 \fi
}%
\providecommand \@ifx [1]{%
 \ifx #1\expandafter \@firstoftwo
 \else \expandafter \@secondoftwo
 \fi
}%
\providecommand \natexlab [1]{#1}%
\providecommand \enquote  [1]{``#1''}%
\providecommand \bibnamefont  [1]{#1}%
\providecommand \bibfnamefont [1]{#1}%
\providecommand \citenamefont [1]{#1}%
\providecommand \href@noop [0]{\@secondoftwo}%
\providecommand \href [0]{\begingroup \@sanitize@url \@href}%
\providecommand \@href[1]{\@@startlink{#1}\@@href}%
\providecommand \@@href[1]{\endgroup#1\@@endlink}%
\providecommand \@sanitize@url [0]{\catcode `\\12\catcode `\$12\catcode
  `\&12\catcode `\#12\catcode `\^12\catcode `\_12\catcode `\%12\relax}%
\providecommand \@@startlink[1]{}%
\providecommand \@@endlink[0]{}%
\providecommand \url  [0]{\begingroup\@sanitize@url \@url }%
\providecommand \@url [1]{\endgroup\@href {#1}{\urlprefix }}%
\providecommand \urlprefix  [0]{URL }%
\providecommand \Eprint [0]{\href }%
\providecommand \doibase [0]{http://dx.doi.org/}%
\providecommand \selectlanguage [0]{\@gobble}%
\providecommand \bibinfo  [0]{\@secondoftwo}%
\providecommand \bibfield  [0]{\@secondoftwo}%
\providecommand \translation [1]{[#1]}%
\providecommand \BibitemOpen [0]{}%
\providecommand \bibitemStop [0]{}%
\providecommand \bibitemNoStop [0]{.\EOS\space}%
\providecommand \EOS [0]{\spacefactor3000\relax}%
\providecommand \BibitemShut  [1]{\csname bibitem#1\endcsname}%
\let\auto@bib@innerbib\@empty
\bibitem [{\citenamefont {Read}\ and\ \citenamefont {Green}(2000)}]{Read1999}%
  \BibitemOpen
  \bibfield  {author} {\bibinfo {author} {\bibfnamefont {N.}~\bibnamefont
  {Read}}\ and\ \bibinfo {author} {\bibfnamefont {D.}~\bibnamefont {Green}},\
  }\href {\doibase 10.1103/PhysRevB.61.10267} {\bibfield  {journal} {\bibinfo
  {journal} {Physical Review B}\ }\textbf {\bibinfo {volume} {61}},\ \bibinfo
  {pages} {10267} (\bibinfo {year} {2000})}\BibitemShut {NoStop}%
\bibitem [{\citenamefont {{Das Sarma}}\ \emph {et~al.}(2005)\citenamefont {{Das
  Sarma}}, \citenamefont {Freedman},\ and\ \citenamefont
  {Nayak}}]{DasSarma2005}%
  \BibitemOpen
  \bibfield  {author} {\bibinfo {author} {\bibfnamefont {S.}~\bibnamefont {{Das
  Sarma}}}, \bibinfo {author} {\bibfnamefont {M.}~\bibnamefont {Freedman}}, \
  and\ \bibinfo {author} {\bibfnamefont {C.}~\bibnamefont {Nayak}},\ }\href
  {\doibase 10.1103/PhysRevLett.94.166802} {\bibfield  {journal} {\bibinfo
  {journal} {Physical Review Letters}\ }\textbf {\bibinfo {volume} {94}},\
  \bibinfo {pages} {166802} (\bibinfo {year} {2005})}\BibitemShut {NoStop}%
\bibitem [{\citenamefont {Nayak}\ \emph {et~al.}(2008)\citenamefont {Nayak},
  \citenamefont {Simon}, \citenamefont {Stern}, \citenamefont {Freedman},\ and\
  \citenamefont {{Das Sarma}}}]{Nayak2008}%
  \BibitemOpen
  \bibfield  {author} {\bibinfo {author} {\bibfnamefont {C.}~\bibnamefont
  {Nayak}}, \bibinfo {author} {\bibfnamefont {S.~H.}\ \bibnamefont {Simon}},
  \bibinfo {author} {\bibfnamefont {A.}~\bibnamefont {Stern}}, \bibinfo
  {author} {\bibfnamefont {M.}~\bibnamefont {Freedman}}, \ and\ \bibinfo
  {author} {\bibfnamefont {S.}~\bibnamefont {{Das Sarma}}},\ }\href {\doibase
  10.1103/RevModPhys.80.1083} {\bibfield  {journal} {\bibinfo  {journal}
  {Reviews of Modern Physics}\ }\textbf {\bibinfo {volume} {80}},\ \bibinfo
  {pages} {1083} (\bibinfo {year} {2008})}\BibitemShut {NoStop}%
\bibitem [{\citenamefont {Kitaev}(2001)}]{Kitaev2000}%
  \BibitemOpen
  \bibfield  {author} {\bibinfo {author} {\bibfnamefont {A.~Y.}\ \bibnamefont
  {Kitaev}},\ }\href {\doibase 10.1070/1063-7869/44/10S/S29} {\bibfield
  {journal} {\bibinfo  {journal} {Physics-Uspekhi}\ }\textbf {\bibinfo {volume}
  {44}},\ \bibinfo {pages} {131} (\bibinfo {year} {2001})}\BibitemShut
  {NoStop}%
\bibitem [{\citenamefont {Oreg}\ \emph {et~al.}(2010)\citenamefont {Oreg},
  \citenamefont {Refael},\ and\ \citenamefont {von Oppen}}]{Oreg2010}%
  \BibitemOpen
  \bibfield  {author} {\bibinfo {author} {\bibfnamefont {Y.}~\bibnamefont
  {Oreg}}, \bibinfo {author} {\bibfnamefont {G.}~\bibnamefont {Refael}}, \ and\
  \bibinfo {author} {\bibfnamefont {F.}~\bibnamefont {von Oppen}},\ }\href
  {\doibase 10.1103/PhysRevLett.105.177002} {\bibfield  {journal} {\bibinfo
  {journal} {Physical Review Letters}\ }\textbf {\bibinfo {volume} {105}},\
  \bibinfo {pages} {177002} (\bibinfo {year} {2010})}\BibitemShut {NoStop}%
\bibitem [{\citenamefont {Lutchyn}\ \emph {et~al.}(2010)\citenamefont
  {Lutchyn}, \citenamefont {Sau},\ and\ \citenamefont {{Das
  Sarma}}}]{Lutchyn2010}%
  \BibitemOpen
  \bibfield  {author} {\bibinfo {author} {\bibfnamefont {R.~M.}\ \bibnamefont
  {Lutchyn}}, \bibinfo {author} {\bibfnamefont {J.~D.}\ \bibnamefont {Sau}}, \
  and\ \bibinfo {author} {\bibfnamefont {S.}~\bibnamefont {{Das Sarma}}},\
  }\href {\doibase 10.1103/PhysRevLett.105.077001} {\bibfield  {journal}
  {\bibinfo  {journal} {Physical Review Letters}\ }\textbf {\bibinfo {volume}
  {105}},\ \bibinfo {pages} {077001} (\bibinfo {year} {2010})}\BibitemShut
  {NoStop}%
\bibitem [{\citenamefont {Lutchyn}\ \emph {et~al.}(2018)\citenamefont
  {Lutchyn}, \citenamefont {Bakkers}, \citenamefont {Kouwenhoven},
  \citenamefont {Krogstrup}, \citenamefont {Marcus},\ and\ \citenamefont
  {Oreg}}]{Lutchyn2018}%
  \BibitemOpen
  \bibfield  {author} {\bibinfo {author} {\bibfnamefont {R.~M.}\ \bibnamefont
  {Lutchyn}}, \bibinfo {author} {\bibfnamefont {E.~P. A.~M.}\ \bibnamefont
  {Bakkers}}, \bibinfo {author} {\bibfnamefont {L.~P.}\ \bibnamefont
  {Kouwenhoven}}, \bibinfo {author} {\bibfnamefont {P.}~\bibnamefont
  {Krogstrup}}, \bibinfo {author} {\bibfnamefont {C.~M.}\ \bibnamefont
  {Marcus}}, \ and\ \bibinfo {author} {\bibfnamefont {Y.}~\bibnamefont
  {Oreg}},\ }\href {\doibase 10.1038/s41578-018-0003-1} {\bibfield  {journal}
  {\bibinfo  {journal} {Nature Reviews Materials}\ }\textbf {\bibinfo {volume}
  {3}},\ \bibinfo {pages} {52} (\bibinfo {year} {2018})}\BibitemShut {NoStop}%
\bibitem [{\citenamefont {Das}\ \emph {et~al.}(2012)\citenamefont {Das},
  \citenamefont {Ronen}, \citenamefont {Most}, \citenamefont {Oreg},
  \citenamefont {Heiblum},\ and\ \citenamefont {Shtrikman}}]{Das2012}%
  \BibitemOpen
  \bibfield  {author} {\bibinfo {author} {\bibfnamefont {A.}~\bibnamefont
  {Das}}, \bibinfo {author} {\bibfnamefont {Y.}~\bibnamefont {Ronen}}, \bibinfo
  {author} {\bibfnamefont {Y.}~\bibnamefont {Most}}, \bibinfo {author}
  {\bibfnamefont {Y.}~\bibnamefont {Oreg}}, \bibinfo {author} {\bibfnamefont
  {M.}~\bibnamefont {Heiblum}}, \ and\ \bibinfo {author} {\bibfnamefont
  {H.}~\bibnamefont {Shtrikman}},\ }\href {\doibase 10.1038/nphys2479}
  {\bibfield  {journal} {\bibinfo  {journal} {Nature Physics}\ }\textbf
  {\bibinfo {volume} {8}},\ \bibinfo {pages} {887} (\bibinfo {year}
  {2012})}\BibitemShut {NoStop}%
\bibitem [{\citenamefont {Mourik}\ \emph {et~al.}(2012)\citenamefont {Mourik},
  \citenamefont {Zuo}, \citenamefont {Frolov}, \citenamefont {Plissard},
  \citenamefont {Bakkers},\ and\ \citenamefont {Kouwenhoven}}]{Mourik2012b}%
  \BibitemOpen
  \bibfield  {author} {\bibinfo {author} {\bibfnamefont {V.}~\bibnamefont
  {Mourik}}, \bibinfo {author} {\bibfnamefont {K.}~\bibnamefont {Zuo}},
  \bibinfo {author} {\bibfnamefont {S.~M.}\ \bibnamefont {Frolov}}, \bibinfo
  {author} {\bibfnamefont {S.~R.}\ \bibnamefont {Plissard}}, \bibinfo {author}
  {\bibfnamefont {E.~P. A.~M.}\ \bibnamefont {Bakkers}}, \ and\ \bibinfo
  {author} {\bibfnamefont {L.~P.}\ \bibnamefont {Kouwenhoven}},\ }\href
  {\doibase 10.1126/science.1222360} {\bibfield  {journal} {\bibinfo  {journal}
  {Science}\ }\textbf {\bibinfo {volume} {336}},\ \bibinfo {pages} {1003}
  (\bibinfo {year} {2012})}\BibitemShut {NoStop}%
\bibitem [{\citenamefont {Deng}\ \emph {et~al.}(2012)\citenamefont {Deng},
  \citenamefont {Yu}, \citenamefont {Huang}, \citenamefont {Larsson},
  \citenamefont {Caroff},\ and\ \citenamefont {Xu}}]{Deng2012}%
  \BibitemOpen
  \bibfield  {author} {\bibinfo {author} {\bibfnamefont {M.~T.}\ \bibnamefont
  {Deng}}, \bibinfo {author} {\bibfnamefont {C.~L.}\ \bibnamefont {Yu}},
  \bibinfo {author} {\bibfnamefont {G.~Y.}\ \bibnamefont {Huang}}, \bibinfo
  {author} {\bibfnamefont {M.}~\bibnamefont {Larsson}}, \bibinfo {author}
  {\bibfnamefont {P.}~\bibnamefont {Caroff}}, \ and\ \bibinfo {author}
  {\bibfnamefont {H.~Q.}\ \bibnamefont {Xu}},\ }\href {\doibase
  10.1021/nl303758w} {\bibfield  {journal} {\bibinfo  {journal} {Nano Letters}\
  }\textbf {\bibinfo {volume} {12}},\ \bibinfo {pages} {6414} (\bibinfo {year}
  {2012})}\BibitemShut {NoStop}%
\bibitem [{\citenamefont {Lee}\ \emph {et~al.}(2012)\citenamefont {Lee},
  \citenamefont {Jiang}, \citenamefont {Aguado}, \citenamefont {Katsaros},
  \citenamefont {Lieber},\ and\ \citenamefont {{De Franceschi}}}]{Lee2012a}%
  \BibitemOpen
  \bibfield  {author} {\bibinfo {author} {\bibfnamefont {E.~J.~H.}\
  \bibnamefont {Lee}}, \bibinfo {author} {\bibfnamefont {X.}~\bibnamefont
  {Jiang}}, \bibinfo {author} {\bibfnamefont {R.}~\bibnamefont {Aguado}},
  \bibinfo {author} {\bibfnamefont {G.}~\bibnamefont {Katsaros}}, \bibinfo
  {author} {\bibfnamefont {C.~M.}\ \bibnamefont {Lieber}}, \ and\ \bibinfo
  {author} {\bibfnamefont {S.}~\bibnamefont {{De Franceschi}}},\ }\href
  {\doibase 10.1103/PhysRevLett.109.186802} {\bibfield  {journal} {\bibinfo
  {journal} {Physical Review Letters}\ }\textbf {\bibinfo {volume} {109}},\
  \bibinfo {pages} {186802} (\bibinfo {year} {2012})}\BibitemShut {NoStop}%
\bibitem [{\citenamefont {Finck}\ \emph {et~al.}(2013)\citenamefont {Finck},
  \citenamefont {{Van Harlingen}}, \citenamefont {Mohseni}, \citenamefont
  {Jung},\ and\ \citenamefont {Li}}]{Finck2013}%
  \BibitemOpen
  \bibfield  {author} {\bibinfo {author} {\bibfnamefont {A.~D.~K.}\
  \bibnamefont {Finck}}, \bibinfo {author} {\bibfnamefont {D.~J.}\ \bibnamefont
  {{Van Harlingen}}}, \bibinfo {author} {\bibfnamefont {P.~K.}\ \bibnamefont
  {Mohseni}}, \bibinfo {author} {\bibfnamefont {K.}~\bibnamefont {Jung}}, \
  and\ \bibinfo {author} {\bibfnamefont {X.}~\bibnamefont {Li}},\ }\href
  {\doibase 10.1103/PhysRevLett.110.126406} {\bibfield  {journal} {\bibinfo
  {journal} {Physical Review Letters}\ }\textbf {\bibinfo {volume} {110}},\
  \bibinfo {pages} {126406} (\bibinfo {year} {2013})}\BibitemShut {NoStop}%
\bibitem [{\citenamefont {Churchill}\ \emph {et~al.}(2013)\citenamefont
  {Churchill}, \citenamefont {Fatemi}, \citenamefont {Grove-Rasmussen},
  \citenamefont {Deng}, \citenamefont {Caroff}, \citenamefont {Xu},\ and\
  \citenamefont {Marcus}}]{Churchill2013}%
  \BibitemOpen
  \bibfield  {author} {\bibinfo {author} {\bibfnamefont {H.~O.~H.}\
  \bibnamefont {Churchill}}, \bibinfo {author} {\bibfnamefont {V.}~\bibnamefont
  {Fatemi}}, \bibinfo {author} {\bibfnamefont {K.}~\bibnamefont
  {Grove-Rasmussen}}, \bibinfo {author} {\bibfnamefont {M.~T.}\ \bibnamefont
  {Deng}}, \bibinfo {author} {\bibfnamefont {P.}~\bibnamefont {Caroff}},
  \bibinfo {author} {\bibfnamefont {H.~Q.}\ \bibnamefont {Xu}}, \ and\ \bibinfo
  {author} {\bibfnamefont {C.~M.}\ \bibnamefont {Marcus}},\ }\href {\doibase
  10.1103/PhysRevB.87.241401} {\bibfield  {journal} {\bibinfo  {journal}
  {Physical Review B}\ }\textbf {\bibinfo {volume} {87}},\ \bibinfo {pages}
  {241401} (\bibinfo {year} {2013})}\BibitemShut {NoStop}%
\bibitem [{\citenamefont {Deng}\ \emph {et~al.}(2015)\citenamefont {Deng},
  \citenamefont {Yu}, \citenamefont {Huang}, \citenamefont {Larsson},
  \citenamefont {Caroff}, \citenamefont {Xu}, \citenamefont {Caroff},
  \citenamefont {Larsson}, \citenamefont {Huang}, \citenamefont {Deng},
  \citenamefont {Yu}, \citenamefont {Huang}, \citenamefont {Larsson},
  \citenamefont {Caroff},\ and\ \citenamefont {Xu}}]{Deng2015a}%
  \BibitemOpen
  \bibfield  {author} {\bibinfo {author} {\bibfnamefont {M.~T.}\ \bibnamefont
  {Deng}}, \bibinfo {author} {\bibfnamefont {C.~L.}\ \bibnamefont {Yu}},
  \bibinfo {author} {\bibfnamefont {G.~Y.}\ \bibnamefont {Huang}}, \bibinfo
  {author} {\bibfnamefont {M.}~\bibnamefont {Larsson}}, \bibinfo {author}
  {\bibfnamefont {P.}~\bibnamefont {Caroff}}, \bibinfo {author} {\bibfnamefont
  {H.~Q.}\ \bibnamefont {Xu}}, \bibinfo {author} {\bibfnamefont
  {P.}~\bibnamefont {Caroff}}, \bibinfo {author} {\bibfnamefont
  {M.}~\bibnamefont {Larsson}}, \bibinfo {author} {\bibfnamefont {G.~Y.}\
  \bibnamefont {Huang}}, \bibinfo {author} {\bibfnamefont {M.~T.}\ \bibnamefont
  {Deng}}, \bibinfo {author} {\bibfnamefont {C.~L.}\ \bibnamefont {Yu}},
  \bibinfo {author} {\bibfnamefont {G.~Y.}\ \bibnamefont {Huang}}, \bibinfo
  {author} {\bibfnamefont {M.}~\bibnamefont {Larsson}}, \bibinfo {author}
  {\bibfnamefont {P.}~\bibnamefont {Caroff}}, \ and\ \bibinfo {author}
  {\bibfnamefont {H.~Q.}\ \bibnamefont {Xu}},\ }\href {\doibase
  10.1038/srep07261} {\bibfield  {journal} {\bibinfo  {journal} {Scientific
  Reports}\ }\textbf {\bibinfo {volume} {4}},\ \bibinfo {pages} {7261}
  (\bibinfo {year} {2015})}\BibitemShut {NoStop}%
\bibitem [{\citenamefont {Takei}\ \emph {et~al.}(2013)\citenamefont {Takei},
  \citenamefont {Fregoso}, \citenamefont {Hui}, \citenamefont {Lobos},\ and\
  \citenamefont {{Das Sarma}}}]{Takei2013}%
  \BibitemOpen
  \bibfield  {author} {\bibinfo {author} {\bibfnamefont {S.}~\bibnamefont
  {Takei}}, \bibinfo {author} {\bibfnamefont {B.~M.}\ \bibnamefont {Fregoso}},
  \bibinfo {author} {\bibfnamefont {H.~Y.}\ \bibnamefont {Hui}}, \bibinfo
  {author} {\bibfnamefont {A.~M.}\ \bibnamefont {Lobos}}, \ and\ \bibinfo
  {author} {\bibfnamefont {S.}~\bibnamefont {{Das Sarma}}},\ }\href {\doibase
  10.1103/PhysRevLett.110.186803} {\bibfield  {journal} {\bibinfo  {journal}
  {Physical Review Letters}\ }\textbf {\bibinfo {volume} {110}},\ \bibinfo
  {pages} {186803} (\bibinfo {year} {2013})}\BibitemShut {NoStop}%
\bibitem [{\citenamefont {Cole}\ \emph {et~al.}(2015)\citenamefont {Cole},
  \citenamefont {{Das Sarma}},\ and\ \citenamefont {Stanescu}}]{Cole2015}%
  \BibitemOpen
  \bibfield  {author} {\bibinfo {author} {\bibfnamefont {W.~S.}\ \bibnamefont
  {Cole}}, \bibinfo {author} {\bibfnamefont {S.}~\bibnamefont {{Das Sarma}}}, \
  and\ \bibinfo {author} {\bibfnamefont {T.~D.}\ \bibnamefont {Stanescu}},\
  }\href {\doibase 10.1103/PhysRevB.92.174511} {\bibfield  {journal} {\bibinfo
  {journal} {Physical Review B}\ }\textbf {\bibinfo {volume} {92}},\ \bibinfo
  {pages} {174511} (\bibinfo {year} {2015})}\BibitemShut {NoStop}%
\bibitem [{\citenamefont {Rainis}\ and\ \citenamefont
  {Loss}(2012)}]{Rainis2012}%
  \BibitemOpen
  \bibfield  {author} {\bibinfo {author} {\bibfnamefont {D.}~\bibnamefont
  {Rainis}}\ and\ \bibinfo {author} {\bibfnamefont {D.}~\bibnamefont {Loss}},\
  }\href {\doibase 10.1103/PhysRevB.85.174533} {\bibfield  {journal} {\bibinfo
  {journal} {Physical Review B - Condensed Matter and Materials Physics}\
  }\textbf {\bibinfo {volume} {85}},\ \bibinfo {pages} {1} (\bibinfo {year}
  {2012})}\BibitemShut {NoStop}%
\bibitem [{\citenamefont {Higginbotham}\ \emph {et~al.}(2015)\citenamefont
  {Higginbotham}, \citenamefont {Albrecht}, \citenamefont {Kir{\v{s}}anskas},
  \citenamefont {Chang}, \citenamefont {Kuemmeth}, \citenamefont {Krogstrup},
  \citenamefont {Jespersen}, \citenamefont {Nyg{\aa}rd}, \citenamefont
  {Flensberg},\ and\ \citenamefont {Marcus}}]{Higginbotham2015}%
  \BibitemOpen
  \bibfield  {author} {\bibinfo {author} {\bibfnamefont {A.~P.}\ \bibnamefont
  {Higginbotham}}, \bibinfo {author} {\bibfnamefont {S.~M.}\ \bibnamefont
  {Albrecht}}, \bibinfo {author} {\bibfnamefont {G.}~\bibnamefont
  {Kir{\v{s}}anskas}}, \bibinfo {author} {\bibfnamefont {W.}~\bibnamefont
  {Chang}}, \bibinfo {author} {\bibfnamefont {F.}~\bibnamefont {Kuemmeth}},
  \bibinfo {author} {\bibfnamefont {P.}~\bibnamefont {Krogstrup}}, \bibinfo
  {author} {\bibfnamefont {T.~S.}\ \bibnamefont {Jespersen}}, \bibinfo {author}
  {\bibfnamefont {J.}~\bibnamefont {Nyg{\aa}rd}}, \bibinfo {author}
  {\bibfnamefont {K.}~\bibnamefont {Flensberg}}, \ and\ \bibinfo {author}
  {\bibfnamefont {C.~M.}\ \bibnamefont {Marcus}},\ }\href {\doibase
  10.1038/nphys3461} {\bibfield  {journal} {\bibinfo  {journal} {Nature
  Physics}\ }\textbf {\bibinfo {volume} {11}},\ \bibinfo {pages} {1017}
  (\bibinfo {year} {2015})}\BibitemShut {NoStop}%
\bibitem [{\citenamefont {Albrecht}\ \emph {et~al.}(2017)\citenamefont
  {Albrecht}, \citenamefont {Hansen}, \citenamefont {Higginbotham},
  \citenamefont {Kuemmeth}, \citenamefont {Jespersen}, \citenamefont
  {Nyg{\aa}rd}, \citenamefont {Krogstrup}, \citenamefont {Danon}, \citenamefont
  {Flensberg},\ and\ \citenamefont {Marcus}}]{Albrecht2017}%
  \BibitemOpen
  \bibfield  {author} {\bibinfo {author} {\bibfnamefont {S.~M.}\ \bibnamefont
  {Albrecht}}, \bibinfo {author} {\bibfnamefont {E.~B.}\ \bibnamefont
  {Hansen}}, \bibinfo {author} {\bibfnamefont {A.~P.}\ \bibnamefont
  {Higginbotham}}, \bibinfo {author} {\bibfnamefont {F.}~\bibnamefont
  {Kuemmeth}}, \bibinfo {author} {\bibfnamefont {T.~S.}\ \bibnamefont
  {Jespersen}}, \bibinfo {author} {\bibfnamefont {J.}~\bibnamefont
  {Nyg{\aa}rd}}, \bibinfo {author} {\bibfnamefont {P.}~\bibnamefont
  {Krogstrup}}, \bibinfo {author} {\bibfnamefont {J.}~\bibnamefont {Danon}},
  \bibinfo {author} {\bibfnamefont {K.}~\bibnamefont {Flensberg}}, \ and\
  \bibinfo {author} {\bibfnamefont {C.~M.}\ \bibnamefont {Marcus}},\ }\href
  {\doibase 10.1103/PhysRevLett.118.137701} {\bibfield  {journal} {\bibinfo
  {journal} {Physical Review Letters}\ }\textbf {\bibinfo {volume} {118}},\
  \bibinfo {pages} {137701} (\bibinfo {year} {2017})}\BibitemShut {NoStop}%
\bibitem [{\citenamefont {Chang}\ \emph {et~al.}(2015)\citenamefont {Chang},
  \citenamefont {Albrecht}, \citenamefont {Jespersen}, \citenamefont
  {Kuemmeth}, \citenamefont {Krogstrup}, \citenamefont {Nyg{\aa}rd},\ and\
  \citenamefont {Marcus}}]{Chang2015c}%
  \BibitemOpen
  \bibfield  {author} {\bibinfo {author} {\bibfnamefont {W.}~\bibnamefont
  {Chang}}, \bibinfo {author} {\bibfnamefont {S.~M.}\ \bibnamefont {Albrecht}},
  \bibinfo {author} {\bibfnamefont {T.~S.}\ \bibnamefont {Jespersen}}, \bibinfo
  {author} {\bibfnamefont {F.}~\bibnamefont {Kuemmeth}}, \bibinfo {author}
  {\bibfnamefont {P.}~\bibnamefont {Krogstrup}}, \bibinfo {author}
  {\bibfnamefont {J.}~\bibnamefont {Nyg{\aa}rd}}, \ and\ \bibinfo {author}
  {\bibfnamefont {C.~M.}\ \bibnamefont {Marcus}},\ }\href {\doibase
  10.1038/nnano.2014.306} {\bibfield  {journal} {\bibinfo  {journal} {Nature
  Nanotechnology}\ }\textbf {\bibinfo {volume} {10}},\ \bibinfo {pages} {232}
  (\bibinfo {year} {2015})}\BibitemShut {NoStop}%
\bibitem [{\citenamefont {Kjaergaard}\ \emph {et~al.}(2016)\citenamefont
  {Kjaergaard}, \citenamefont {Nichele}, \citenamefont {Suominen},
  \citenamefont {Nowak}, \citenamefont {Wimmer}, \citenamefont {Akhmerov},
  \citenamefont {Folk}, \citenamefont {Flensberg}, \citenamefont {Shabani},
  \citenamefont {Palmstr{\o}m},\ and\ \citenamefont
  {Marcus}}]{Kjaergaard2016a}%
  \BibitemOpen
  \bibfield  {author} {\bibinfo {author} {\bibfnamefont {M.}~\bibnamefont
  {Kjaergaard}}, \bibinfo {author} {\bibfnamefont {F.}~\bibnamefont {Nichele}},
  \bibinfo {author} {\bibfnamefont {H.~J.}\ \bibnamefont {Suominen}}, \bibinfo
  {author} {\bibfnamefont {M.~P.}\ \bibnamefont {Nowak}}, \bibinfo {author}
  {\bibfnamefont {M.}~\bibnamefont {Wimmer}}, \bibinfo {author} {\bibfnamefont
  {A.~R.}\ \bibnamefont {Akhmerov}}, \bibinfo {author} {\bibfnamefont {J.~A.}\
  \bibnamefont {Folk}}, \bibinfo {author} {\bibfnamefont {K.}~\bibnamefont
  {Flensberg}}, \bibinfo {author} {\bibfnamefont {J.}~\bibnamefont {Shabani}},
  \bibinfo {author} {\bibfnamefont {C.~J.}\ \bibnamefont {Palmstr{\o}m}}, \
  and\ \bibinfo {author} {\bibfnamefont {C.~M.}\ \bibnamefont {Marcus}},\
  }\href {\doibase 10.1038/ncomms12841} {\bibfield  {journal} {\bibinfo
  {journal} {Nature Communications}\ }\textbf {\bibinfo {volume} {7}},\
  \bibinfo {pages} {12841} (\bibinfo {year} {2016})}\BibitemShut {NoStop}%
\bibitem [{\citenamefont {Zhang}\ \emph {et~al.}(2017)\citenamefont {Zhang},
  \citenamefont {G{\"{u}}l}, \citenamefont {Conesa-Boj}, \citenamefont {Nowak},
  \citenamefont {Wimmer}, \citenamefont {Zuo}, \citenamefont {Mourik},
  \citenamefont {de~Vries}, \citenamefont {van Veen}, \citenamefont {de~Moor},
  \citenamefont {Bommer}, \citenamefont {van Woerkom}, \citenamefont {Car},
  \citenamefont {Plissard}, \citenamefont {Bakkers}, \citenamefont
  {Quintero-P{\'{e}}rez}, \citenamefont {Cassidy}, \citenamefont {Koelling},
  \citenamefont {Goswami}, \citenamefont {Watanabe}, \citenamefont
  {Taniguchi},\ and\ \citenamefont {Kouwenhoven}}]{Zhang2017}%
  \BibitemOpen
  \bibfield  {author} {\bibinfo {author} {\bibfnamefont {H.}~\bibnamefont
  {Zhang}}, \bibinfo {author} {\bibfnamefont {{\"{O}}.}~\bibnamefont
  {G{\"{u}}l}}, \bibinfo {author} {\bibfnamefont {S.}~\bibnamefont
  {Conesa-Boj}}, \bibinfo {author} {\bibfnamefont {M.~P.}\ \bibnamefont
  {Nowak}}, \bibinfo {author} {\bibfnamefont {M.}~\bibnamefont {Wimmer}},
  \bibinfo {author} {\bibfnamefont {K.}~\bibnamefont {Zuo}}, \bibinfo {author}
  {\bibfnamefont {V.}~\bibnamefont {Mourik}}, \bibinfo {author} {\bibfnamefont
  {F.~K.}\ \bibnamefont {de~Vries}}, \bibinfo {author} {\bibfnamefont
  {J.}~\bibnamefont {van Veen}}, \bibinfo {author} {\bibfnamefont {M.~W.~A.}\
  \bibnamefont {de~Moor}}, \bibinfo {author} {\bibfnamefont {J.~D.~S.}\
  \bibnamefont {Bommer}}, \bibinfo {author} {\bibfnamefont {D.~J.}\
  \bibnamefont {van Woerkom}}, \bibinfo {author} {\bibfnamefont
  {D.}~\bibnamefont {Car}}, \bibinfo {author} {\bibfnamefont {S.~R.}\
  \bibnamefont {Plissard}}, \bibinfo {author} {\bibfnamefont {E.~P. A.~M.}\
  \bibnamefont {Bakkers}}, \bibinfo {author} {\bibfnamefont {M.}~\bibnamefont
  {Quintero-P{\'{e}}rez}}, \bibinfo {author} {\bibfnamefont {M.~C.}\
  \bibnamefont {Cassidy}}, \bibinfo {author} {\bibfnamefont {S.}~\bibnamefont
  {Koelling}}, \bibinfo {author} {\bibfnamefont {S.}~\bibnamefont {Goswami}},
  \bibinfo {author} {\bibfnamefont {K.}~\bibnamefont {Watanabe}}, \bibinfo
  {author} {\bibfnamefont {T.}~\bibnamefont {Taniguchi}}, \ and\ \bibinfo
  {author} {\bibfnamefont {L.~P.}\ \bibnamefont {Kouwenhoven}},\ }\href
  {\doibase 10.1038/ncomms16025} {\bibfield  {journal} {\bibinfo  {journal}
  {Nature Communications}\ }\textbf {\bibinfo {volume} {8}},\ \bibinfo {pages}
  {16025} (\bibinfo {year} {2017})}\BibitemShut {NoStop}%
\bibitem [{\citenamefont {G{\"{u}}l}\ \emph {et~al.}(2017)\citenamefont
  {G{\"{u}}l}, \citenamefont {Zhang}, \citenamefont {de~Vries}, \citenamefont
  {van Veen}, \citenamefont {Zuo}, \citenamefont {Mourik}, \citenamefont
  {Conesa-Boj}, \citenamefont {Nowak}, \citenamefont {van Woerkom},
  \citenamefont {Quintero-P{\'{e}}rez}, \citenamefont {Cassidy}, \citenamefont
  {Geresdi}, \citenamefont {Koelling}, \citenamefont {Car}, \citenamefont
  {Plissard}, \citenamefont {Bakkers},\ and\ \citenamefont
  {Kouwenhoven}}]{Gul2017}%
  \BibitemOpen
  \bibfield  {author} {\bibinfo {author} {\bibfnamefont {{\"{O}}.}~\bibnamefont
  {G{\"{u}}l}}, \bibinfo {author} {\bibfnamefont {H.}~\bibnamefont {Zhang}},
  \bibinfo {author} {\bibfnamefont {F.~K.}\ \bibnamefont {de~Vries}}, \bibinfo
  {author} {\bibfnamefont {J.}~\bibnamefont {van Veen}}, \bibinfo {author}
  {\bibfnamefont {K.}~\bibnamefont {Zuo}}, \bibinfo {author} {\bibfnamefont
  {V.}~\bibnamefont {Mourik}}, \bibinfo {author} {\bibfnamefont
  {S.}~\bibnamefont {Conesa-Boj}}, \bibinfo {author} {\bibfnamefont {M.~P.}\
  \bibnamefont {Nowak}}, \bibinfo {author} {\bibfnamefont {D.~J.}\ \bibnamefont
  {van Woerkom}}, \bibinfo {author} {\bibfnamefont {M.}~\bibnamefont
  {Quintero-P{\'{e}}rez}}, \bibinfo {author} {\bibfnamefont {M.~C.}\
  \bibnamefont {Cassidy}}, \bibinfo {author} {\bibfnamefont {A.}~\bibnamefont
  {Geresdi}}, \bibinfo {author} {\bibfnamefont {S.}~\bibnamefont {Koelling}},
  \bibinfo {author} {\bibfnamefont {D.}~\bibnamefont {Car}}, \bibinfo {author}
  {\bibfnamefont {S.~R.}\ \bibnamefont {Plissard}}, \bibinfo {author}
  {\bibfnamefont {E.~P. A.~M.}\ \bibnamefont {Bakkers}}, \ and\ \bibinfo
  {author} {\bibfnamefont {L.~P.}\ \bibnamefont {Kouwenhoven}},\ }\href
  {\doibase 10.1021/acs.nanolett.7b00540} {\bibfield  {journal} {\bibinfo
  {journal} {Nano Letters}\ }\textbf {\bibinfo {volume} {17}},\ \bibinfo
  {pages} {2690} (\bibinfo {year} {2017})}\BibitemShut {NoStop}%
\bibitem [{\citenamefont {Albrecht}\ \emph {et~al.}(2016)\citenamefont
  {Albrecht}, \citenamefont {Higginbotham}, \citenamefont {Madsen},
  \citenamefont {Kuemmeth}, \citenamefont {Jespersen}, \citenamefont
  {Nyg{\aa}rd}, \citenamefont {Krogstrup},\ and\ \citenamefont
  {Marcus}}]{Albrecht2016}%
  \BibitemOpen
  \bibfield  {author} {\bibinfo {author} {\bibfnamefont {S.~M.}\ \bibnamefont
  {Albrecht}}, \bibinfo {author} {\bibfnamefont {A.~P.}\ \bibnamefont
  {Higginbotham}}, \bibinfo {author} {\bibfnamefont {M.}~\bibnamefont
  {Madsen}}, \bibinfo {author} {\bibfnamefont {F.}~\bibnamefont {Kuemmeth}},
  \bibinfo {author} {\bibfnamefont {T.~S.}\ \bibnamefont {Jespersen}}, \bibinfo
  {author} {\bibfnamefont {J.}~\bibnamefont {Nyg{\aa}rd}}, \bibinfo {author}
  {\bibfnamefont {P.}~\bibnamefont {Krogstrup}}, \ and\ \bibinfo {author}
  {\bibfnamefont {C.~M.}\ \bibnamefont {Marcus}},\ }\href {\doibase
  10.1038/nature17162} {\bibfield  {journal} {\bibinfo  {journal} {Nature}\
  }\textbf {\bibinfo {volume} {531}},\ \bibinfo {pages} {206} (\bibinfo {year}
  {2016})}\BibitemShut {NoStop}%
\bibitem [{\citenamefont {Deng}\ \emph {et~al.}(2016)\citenamefont {Deng},
  \citenamefont {Vaitiekėnas}, \citenamefont {Hansen}, \citenamefont {Danon},
  \citenamefont {Leijnse}, \citenamefont {Flensberg}, \citenamefont
  {Nyg{\aa}rd}, \citenamefont {Krogstrup},\ and\ \citenamefont
  {Marcus}}]{Deng2016}%
  \BibitemOpen
  \bibfield  {author} {\bibinfo {author} {\bibfnamefont {M.~T.}\ \bibnamefont
  {Deng}}, \bibinfo {author} {\bibfnamefont {S.}~\bibnamefont {Vaitiekėnas}},
  \bibinfo {author} {\bibfnamefont {E.~B.}\ \bibnamefont {Hansen}}, \bibinfo
  {author} {\bibfnamefont {J.}~\bibnamefont {Danon}}, \bibinfo {author}
  {\bibfnamefont {M.}~\bibnamefont {Leijnse}}, \bibinfo {author} {\bibfnamefont
  {K.}~\bibnamefont {Flensberg}}, \bibinfo {author} {\bibfnamefont
  {J.}~\bibnamefont {Nyg{\aa}rd}}, \bibinfo {author} {\bibfnamefont
  {P.}~\bibnamefont {Krogstrup}}, \ and\ \bibinfo {author} {\bibfnamefont
  {C.~M.}\ \bibnamefont {Marcus}},\ }\href {\doibase 10.1126/science.aaf3961}
  {\bibfield  {journal} {\bibinfo  {journal} {Science}\ }\textbf {\bibinfo
  {volume} {354}},\ \bibinfo {pages} {1557} (\bibinfo {year}
  {2016})}\BibitemShut {NoStop}%
\bibitem [{\citenamefont {G{\"{u}}l}\ \emph {et~al.}(2018)\citenamefont
  {G{\"{u}}l}, \citenamefont {Zhang}, \citenamefont {Bommer}, \citenamefont
  {de~Moor}, \citenamefont {Car}, \citenamefont {Plissard}, \citenamefont
  {Bakkers}, \citenamefont {Geresdi}, \citenamefont {Watanabe}, \citenamefont
  {Taniguchi},\ and\ \citenamefont {Kouwenhoven}}]{Gul2018}%
  \BibitemOpen
  \bibfield  {author} {\bibinfo {author} {\bibfnamefont {{\"{O}}.}~\bibnamefont
  {G{\"{u}}l}}, \bibinfo {author} {\bibfnamefont {H.}~\bibnamefont {Zhang}},
  \bibinfo {author} {\bibfnamefont {J.~D.~S.}\ \bibnamefont {Bommer}}, \bibinfo
  {author} {\bibfnamefont {M.~W.~A.}\ \bibnamefont {de~Moor}}, \bibinfo
  {author} {\bibfnamefont {D.}~\bibnamefont {Car}}, \bibinfo {author}
  {\bibfnamefont {S.~R.}\ \bibnamefont {Plissard}}, \bibinfo {author}
  {\bibfnamefont {E.~P. A.~M.}\ \bibnamefont {Bakkers}}, \bibinfo {author}
  {\bibfnamefont {A.}~\bibnamefont {Geresdi}}, \bibinfo {author} {\bibfnamefont
  {K.}~\bibnamefont {Watanabe}}, \bibinfo {author} {\bibfnamefont
  {T.}~\bibnamefont {Taniguchi}}, \ and\ \bibinfo {author} {\bibfnamefont
  {L.~P.}\ \bibnamefont {Kouwenhoven}},\ }\href {\doibase
  10.1038/s41565-017-0032-8} {\bibfield  {journal} {\bibinfo  {journal} {Nature
  Nanotechnology}\ }\textbf {\bibinfo {volume} {13}},\ \bibinfo {pages} {192}
  (\bibinfo {year} {2018})}\BibitemShut {NoStop}%
\bibitem [{\citenamefont {Conesa-Boj}\ \emph {et~al.}(2017)\citenamefont
  {Conesa-Boj}, \citenamefont {Li}, \citenamefont {Koelling}, \citenamefont
  {Brauns}, \citenamefont {Ridderbos}, \citenamefont {Nguyen}, \citenamefont
  {Verheijen}, \citenamefont {Koenraad}, \citenamefont {Zwanenburg},\ and\
  \citenamefont {Bakkers}}]{Conesa-Boj2017}%
  \BibitemOpen
  \bibfield  {author} {\bibinfo {author} {\bibfnamefont {S.}~\bibnamefont
  {Conesa-Boj}}, \bibinfo {author} {\bibfnamefont {A.}~\bibnamefont {Li}},
  \bibinfo {author} {\bibfnamefont {S.}~\bibnamefont {Koelling}}, \bibinfo
  {author} {\bibfnamefont {M.}~\bibnamefont {Brauns}}, \bibinfo {author}
  {\bibfnamefont {J.}~\bibnamefont {Ridderbos}}, \bibinfo {author}
  {\bibfnamefont {T.~T.}\ \bibnamefont {Nguyen}}, \bibinfo {author}
  {\bibfnamefont {M.~A.}\ \bibnamefont {Verheijen}}, \bibinfo {author}
  {\bibfnamefont {P.~M.}\ \bibnamefont {Koenraad}}, \bibinfo {author}
  {\bibfnamefont {F.~A.}\ \bibnamefont {Zwanenburg}}, \ and\ \bibinfo {author}
  {\bibfnamefont {E.~P. A.~M.}\ \bibnamefont {Bakkers}},\ }\href {\doibase
  10.1021/acs.nanolett.6b04891} {\bibfield  {journal} {\bibinfo  {journal}
  {Nano Letters}\ }\textbf {\bibinfo {volume} {17}},\ \bibinfo {pages} {2259}
  (\bibinfo {year} {2017})}\BibitemShut {NoStop}%
\bibitem [{\citenamefont {Lu}\ \emph {et~al.}(2005)\citenamefont {Lu},
  \citenamefont {Xiang}, \citenamefont {Timko}, \citenamefont {Wu},\ and\
  \citenamefont {Lieber}}]{Lu2005a}%
  \BibitemOpen
  \bibfield  {author} {\bibinfo {author} {\bibfnamefont {W.}~\bibnamefont
  {Lu}}, \bibinfo {author} {\bibfnamefont {J.}~\bibnamefont {Xiang}}, \bibinfo
  {author} {\bibfnamefont {B.~P.}\ \bibnamefont {Timko}}, \bibinfo {author}
  {\bibfnamefont {Y.}~\bibnamefont {Wu}}, \ and\ \bibinfo {author}
  {\bibfnamefont {C.~M.}\ \bibnamefont {Lieber}},\ }\href {\doibase
  10.1073/pnas.0504581102} {\bibfield  {journal} {\bibinfo  {journal}
  {Proceedings of the National Academy of Sciences}\ }\textbf {\bibinfo
  {volume} {102}},\ \bibinfo {pages} {10046} (\bibinfo {year}
  {2005})}\BibitemShut {NoStop}%
\bibitem [{\citenamefont {Zhang}\ \emph {et~al.}(2010)\citenamefont {Zhang},
  \citenamefont {Lopez}, \citenamefont {Hyun},\ and\ \citenamefont
  {Lauhon}}]{Zhang2010}%
  \BibitemOpen
  \bibfield  {author} {\bibinfo {author} {\bibfnamefont {S.}~\bibnamefont
  {Zhang}}, \bibinfo {author} {\bibfnamefont {F.~J.}\ \bibnamefont {Lopez}},
  \bibinfo {author} {\bibfnamefont {J.~K.}\ \bibnamefont {Hyun}}, \ and\
  \bibinfo {author} {\bibfnamefont {L.~J.}\ \bibnamefont {Lauhon}},\ }\href
  {\doibase 10.1021/nl102316b} {\bibfield  {journal} {\bibinfo  {journal} {Nano
  letters}\ }\textbf {\bibinfo {volume} {10}},\ \bibinfo {pages} {4483}
  (\bibinfo {year} {2010})}\BibitemShut {NoStop}%
\bibitem [{\citenamefont {Maier}\ \emph
  {et~al.}(2014{\natexlab{a}})\citenamefont {Maier}, \citenamefont
  {Klinovaja},\ and\ \citenamefont {Loss}}]{Maier2014}%
  \BibitemOpen
  \bibfield  {author} {\bibinfo {author} {\bibfnamefont {F.}~\bibnamefont
  {Maier}}, \bibinfo {author} {\bibfnamefont {J.}~\bibnamefont {Klinovaja}}, \
  and\ \bibinfo {author} {\bibfnamefont {D.}~\bibnamefont {Loss}},\ }\href
  {\doibase 10.1103/PhysRevB.90.195421} {\bibfield  {journal} {\bibinfo
  {journal} {Physical Review B}\ }\textbf {\bibinfo {volume} {90}},\ \bibinfo
  {pages} {195421} (\bibinfo {year} {2014}{\natexlab{a}})}\BibitemShut
  {NoStop}%
\bibitem [{\citenamefont {Thakurathi}\ \emph {et~al.}(2017)\citenamefont
  {Thakurathi}, \citenamefont {Loss},\ and\ \citenamefont
  {Klinovaja}}]{Thakurathi2017}%
  \BibitemOpen
  \bibfield  {author} {\bibinfo {author} {\bibfnamefont {M.}~\bibnamefont
  {Thakurathi}}, \bibinfo {author} {\bibfnamefont {D.}~\bibnamefont {Loss}}, \
  and\ \bibinfo {author} {\bibfnamefont {J.}~\bibnamefont {Klinovaja}},\ }\href
  {\doibase 10.1103/PhysRevB.95.155407} {\bibfield  {journal} {\bibinfo
  {journal} {Physical Review B}\ }\textbf {\bibinfo {volume} {95}},\ \bibinfo
  {pages} {155407} (\bibinfo {year} {2017})}\BibitemShut {NoStop}%
\bibitem [{\citenamefont {Xiang}\ \emph {et~al.}(2006)\citenamefont {Xiang},
  \citenamefont {Vidan}, \citenamefont {Tinkham}, \citenamefont {Westervelt},\
  and\ \citenamefont {Lieber}}]{Xiang2006b}%
  \BibitemOpen
  \bibfield  {author} {\bibinfo {author} {\bibfnamefont {J.}~\bibnamefont
  {Xiang}}, \bibinfo {author} {\bibfnamefont {A.}~\bibnamefont {Vidan}},
  \bibinfo {author} {\bibfnamefont {M.}~\bibnamefont {Tinkham}}, \bibinfo
  {author} {\bibfnamefont {R.~M.}\ \bibnamefont {Westervelt}}, \ and\ \bibinfo
  {author} {\bibfnamefont {C.~M.}\ \bibnamefont {Lieber}},\ }\href {\doibase
  10.1038/nnano.2006.140} {\bibfield  {journal} {\bibinfo  {journal} {Nature
  Nanotechnology}\ }\textbf {\bibinfo {volume} {1}},\ \bibinfo {pages} {208}
  (\bibinfo {year} {2006})}\BibitemShut {NoStop}%
\bibitem [{\citenamefont {Su}\ \emph {et~al.}(2016)\citenamefont {Su},
  \citenamefont {Zarassi}, \citenamefont {Nguyen}, \citenamefont {Yoo},
  \citenamefont {Dayeh},\ and\ \citenamefont {Frolov}}]{Su2016}%
  \BibitemOpen
  \bibfield  {author} {\bibinfo {author} {\bibfnamefont {Z.}~\bibnamefont
  {Su}}, \bibinfo {author} {\bibfnamefont {A.}~\bibnamefont {Zarassi}},
  \bibinfo {author} {\bibfnamefont {B.-M.}\ \bibnamefont {Nguyen}}, \bibinfo
  {author} {\bibfnamefont {J.}~\bibnamefont {Yoo}}, \bibinfo {author}
  {\bibfnamefont {S.~A.}\ \bibnamefont {Dayeh}}, \ and\ \bibinfo {author}
  {\bibfnamefont {S.~M.}\ \bibnamefont {Frolov}},\ }\href
  {http://arxiv.org/abs/1610.03010} {\bibfield  {journal} {\bibinfo  {journal}
  {arXiv:1610.03010}\ } (\bibinfo {year} {2016})}\BibitemShut {NoStop}%
\bibitem [{\citenamefont {Ridderbos}\ \emph {et~al.}(2018)\citenamefont
  {Ridderbos}, \citenamefont {Brauns}, \citenamefont {Shen}, \citenamefont
  {de~Vries}, \citenamefont {Li}, \citenamefont {Bakkers}, \citenamefont
  {Brinkman},\ and\ \citenamefont {Zwanenburg}}]{Ridderbos2017}%
  \BibitemOpen
  \bibfield  {author} {\bibinfo {author} {\bibfnamefont {J.}~\bibnamefont
  {Ridderbos}}, \bibinfo {author} {\bibfnamefont {M.}~\bibnamefont {Brauns}},
  \bibinfo {author} {\bibfnamefont {J.}~\bibnamefont {Shen}}, \bibinfo {author}
  {\bibfnamefont {F.~K.}\ \bibnamefont {de~Vries}}, \bibinfo {author}
  {\bibfnamefont {A.}~\bibnamefont {Li}}, \bibinfo {author} {\bibfnamefont
  {E.~P. A.~M.}\ \bibnamefont {Bakkers}}, \bibinfo {author} {\bibfnamefont
  {A.}~\bibnamefont {Brinkman}}, \ and\ \bibinfo {author} {\bibfnamefont
  {F.~A.}\ \bibnamefont {Zwanenburg}},\ }\href {\doibase
  10.1002/adma.201802257} {\bibfield  {journal} {\bibinfo  {journal} {Advanced
  Materials}\ }\textbf {\bibinfo {volume} {30}},\ \bibinfo {pages} {1802257}
  (\bibinfo {year} {2018})}\BibitemShut {NoStop}%
\bibitem [{\citenamefont {de~Vries}\ \emph {et~al.}(2018)\citenamefont
  {de~Vries}, \citenamefont {Shen}, \citenamefont {Skolasinski}, \citenamefont
  {Nowak}, \citenamefont {Varjas}, \citenamefont {Wang}, \citenamefont
  {Wimmer}, \citenamefont {Ridderbos}, \citenamefont {Zwanenburg},
  \citenamefont {Li}, \citenamefont {Koelling}, \citenamefont {Verheijen},
  \citenamefont {Bakkers},\ and\ \citenamefont {Kouwenhoven}}]{DeVries2018}%
  \BibitemOpen
  \bibfield  {author} {\bibinfo {author} {\bibfnamefont {F.~K.}\ \bibnamefont
  {de~Vries}}, \bibinfo {author} {\bibfnamefont {J.}~\bibnamefont {Shen}},
  \bibinfo {author} {\bibfnamefont {R.~J.}\ \bibnamefont {Skolasinski}},
  \bibinfo {author} {\bibfnamefont {M.~P.}\ \bibnamefont {Nowak}}, \bibinfo
  {author} {\bibfnamefont {D.}~\bibnamefont {Varjas}}, \bibinfo {author}
  {\bibfnamefont {L.}~\bibnamefont {Wang}}, \bibinfo {author} {\bibfnamefont
  {M.}~\bibnamefont {Wimmer}}, \bibinfo {author} {\bibfnamefont
  {J.}~\bibnamefont {Ridderbos}}, \bibinfo {author} {\bibfnamefont {F.~A.}\
  \bibnamefont {Zwanenburg}}, \bibinfo {author} {\bibfnamefont
  {A.}~\bibnamefont {Li}}, \bibinfo {author} {\bibfnamefont {S.}~\bibnamefont
  {Koelling}}, \bibinfo {author} {\bibfnamefont {M.~A.}\ \bibnamefont
  {Verheijen}}, \bibinfo {author} {\bibfnamefont {E.~P. A.~M.}\ \bibnamefont
  {Bakkers}}, \ and\ \bibinfo {author} {\bibfnamefont {L.~P.}\ \bibnamefont
  {Kouwenhoven}},\ }\href {\doibase 10.1021/acs.nanolett.8b02981} {\bibfield
  {journal} {\bibinfo  {journal} {Nano Letters}\ }\textbf {\bibinfo {volume}
  {18}},\ \bibinfo {pages} {6483} (\bibinfo {year} {2018})}\BibitemShut
  {NoStop}%
\bibitem [{\citenamefont {Ridderbos}\ \emph {et~al.}(2019)\citenamefont
  {Ridderbos}, \citenamefont {Brauns}, \citenamefont {Li}, \citenamefont
  {Bakkers}, \citenamefont {Brinkman}, \citenamefont {van~der Wiel},\ and\
  \citenamefont {Zwanenburg}}]{Ridderbos2019}%
  \BibitemOpen
  \bibfield  {author} {\bibinfo {author} {\bibfnamefont {J.}~\bibnamefont
  {Ridderbos}}, \bibinfo {author} {\bibfnamefont {M.}~\bibnamefont {Brauns}},
  \bibinfo {author} {\bibfnamefont {A.}~\bibnamefont {Li}}, \bibinfo {author}
  {\bibfnamefont {E.~P. A.~M.}\ \bibnamefont {Bakkers}}, \bibinfo {author}
  {\bibfnamefont {A.}~\bibnamefont {Brinkman}}, \bibinfo {author}
  {\bibfnamefont {W.~G.}\ \bibnamefont {van~der Wiel}}, \ and\ \bibinfo
  {author} {\bibfnamefont {F.~A.}\ \bibnamefont {Zwanenburg}},\ }\href
  {\doibase 10.1103/PhysRevMaterials.3.084803} {\bibfield  {journal} {\bibinfo
  {journal} {Physical Review Materials}\ }\textbf {\bibinfo {volume} {3}},\
  \bibinfo {pages} {084803} (\bibinfo {year} {2019})}\BibitemShut {NoStop}%
\bibitem [{\citenamefont {Kloeffel}\ \emph {et~al.}(2011)\citenamefont
  {Kloeffel}, \citenamefont {Trif},\ and\ \citenamefont {Loss}}]{Kloeffel2011}%
  \BibitemOpen
  \bibfield  {author} {\bibinfo {author} {\bibfnamefont {C.}~\bibnamefont
  {Kloeffel}}, \bibinfo {author} {\bibfnamefont {M.}~\bibnamefont {Trif}}, \
  and\ \bibinfo {author} {\bibfnamefont {D.}~\bibnamefont {Loss}},\ }\href
  {\doibase 10.1103/PhysRevB.84.195314} {\bibfield  {journal} {\bibinfo
  {journal} {Physical Review B}\ }\textbf {\bibinfo {volume} {84}},\ \bibinfo
  {pages} {195314} (\bibinfo {year} {2011})}\BibitemShut {NoStop}%
\bibitem [{\citenamefont {Maier}\ \emph {et~al.}(2013)\citenamefont {Maier},
  \citenamefont {Kloeffel},\ and\ \citenamefont {Loss}}]{Maier2013}%
  \BibitemOpen
  \bibfield  {author} {\bibinfo {author} {\bibfnamefont {F.}~\bibnamefont
  {Maier}}, \bibinfo {author} {\bibfnamefont {C.}~\bibnamefont {Kloeffel}}, \
  and\ \bibinfo {author} {\bibfnamefont {D.}~\bibnamefont {Loss}},\ }\href
  {\doibase 10.1103/PhysRevB.87.161305} {\bibfield  {journal} {\bibinfo
  {journal} {Physical Review B}\ }\textbf {\bibinfo {volume} {87}},\ \bibinfo
  {pages} {161305} (\bibinfo {year} {2013})}\BibitemShut {NoStop}%
\bibitem [{\citenamefont {Brauns}\ \emph {et~al.}(2016)\citenamefont {Brauns},
  \citenamefont {Ridderbos}, \citenamefont {Li}, \citenamefont {Bakkers},\ and\
  \citenamefont {Zwanenburg}}]{Brauns2015}%
  \BibitemOpen
  \bibfield  {author} {\bibinfo {author} {\bibfnamefont {M.}~\bibnamefont
  {Brauns}}, \bibinfo {author} {\bibfnamefont {J.}~\bibnamefont {Ridderbos}},
  \bibinfo {author} {\bibfnamefont {A.}~\bibnamefont {Li}}, \bibinfo {author}
  {\bibfnamefont {E.~P. A.~M.}\ \bibnamefont {Bakkers}}, \ and\ \bibinfo
  {author} {\bibfnamefont {F.~A.}\ \bibnamefont {Zwanenburg}},\ }\href
  {\doibase 10.1103/PhysRevB.93.121408} {\bibfield  {journal} {\bibinfo
  {journal} {Physical Review B}\ }\textbf {\bibinfo {volume} {93}},\ \bibinfo
  {pages} {121408(R)} (\bibinfo {year} {2016})}\BibitemShut {NoStop}%
\bibitem [{\citenamefont {Kral}\ \emph {et~al.}(2015)\citenamefont {Kral},
  \citenamefont {Zeiner}, \citenamefont {St{\"{o}}ger-Pollach}, \citenamefont
  {Bertagnolli}, \citenamefont {den Hertog}, \citenamefont {Lopez-Haro},
  \citenamefont {Robin}, \citenamefont {{El Hajraoui}},\ and\ \citenamefont
  {Lugstein}}]{Kral2015}%
  \BibitemOpen
  \bibfield  {author} {\bibinfo {author} {\bibfnamefont {S.}~\bibnamefont
  {Kral}}, \bibinfo {author} {\bibfnamefont {C.}~\bibnamefont {Zeiner}},
  \bibinfo {author} {\bibfnamefont {M.}~\bibnamefont {St{\"{o}}ger-Pollach}},
  \bibinfo {author} {\bibfnamefont {E.}~\bibnamefont {Bertagnolli}}, \bibinfo
  {author} {\bibfnamefont {M.~I.}\ \bibnamefont {den Hertog}}, \bibinfo
  {author} {\bibfnamefont {M.}~\bibnamefont {Lopez-Haro}}, \bibinfo {author}
  {\bibfnamefont {E.}~\bibnamefont {Robin}}, \bibinfo {author} {\bibfnamefont
  {K.}~\bibnamefont {{El Hajraoui}}}, \ and\ \bibinfo {author} {\bibfnamefont
  {A.}~\bibnamefont {Lugstein}},\ }\href {\doibase
  10.1021/acs.nanolett.5b01748} {\bibfield  {journal} {\bibinfo  {journal}
  {Nano Letters}\ }\textbf {\bibinfo {volume} {15}},\ \bibinfo {pages} {4783}
  (\bibinfo {year} {2015})}\BibitemShut {NoStop}%
\bibitem [{\citenamefont {{El Hajraoui}}\ \emph {et~al.}(2019)\citenamefont
  {{El Hajraoui}}, \citenamefont {Luong}, \citenamefont {Robin}, \citenamefont
  {Brunbauer}, \citenamefont {Zeiner}, \citenamefont {Lugstein}, \citenamefont
  {Gentile}, \citenamefont {Rouvi{\`{e}}re},\ and\ \citenamefont {{Den
  Hertog}}}]{ElHajraoui2019}%
  \BibitemOpen
  \bibfield  {author} {\bibinfo {author} {\bibfnamefont {K.}~\bibnamefont {{El
  Hajraoui}}}, \bibinfo {author} {\bibfnamefont {M.~A.}\ \bibnamefont {Luong}},
  \bibinfo {author} {\bibfnamefont {E.}~\bibnamefont {Robin}}, \bibinfo
  {author} {\bibfnamefont {F.}~\bibnamefont {Brunbauer}}, \bibinfo {author}
  {\bibfnamefont {C.}~\bibnamefont {Zeiner}}, \bibinfo {author} {\bibfnamefont
  {A.}~\bibnamefont {Lugstein}}, \bibinfo {author} {\bibfnamefont
  {P.}~\bibnamefont {Gentile}}, \bibinfo {author} {\bibfnamefont {J.~L.}\
  \bibnamefont {Rouvi{\`{e}}re}}, \ and\ \bibinfo {author} {\bibfnamefont
  {M.}~\bibnamefont {{Den Hertog}}},\ }\href {\doibase
  10.1021/acs.nanolett.8b05171} {\bibfield  {journal} {\bibinfo  {journal}
  {Nano Letters}\ }\textbf {\bibinfo {volume} {19}},\ \bibinfo {pages} {2897}
  (\bibinfo {year} {2019})}\BibitemShut {NoStop}%
\bibitem [{\citenamefont {Gale}\ and\ \citenamefont
  {Totemeier}(2003)}]{Gale2003}%
  \BibitemOpen
  \bibfield  {author} {\bibinfo {author} {\bibfnamefont {W.~F.}\ \bibnamefont
  {Gale}}\ and\ \bibinfo {author} {\bibfnamefont {T.~C.}\ \bibnamefont
  {Totemeier}},\ }\href@noop {} {\emph {\bibinfo {title} {{Smithells Metals
  Reference Book 8th edition}}}},\ edited by\ \bibinfo {editor} {\bibfnamefont
  {W.~F.}\ \bibnamefont {Gale.}}\ and\ \bibinfo {editor} {\bibfnamefont
  {T.~C.}\ \bibnamefont {Totemeier}}\ (\bibinfo  {publisher} {Elsevier Ltd},\
  \bibinfo {year} {2003})\ p.\ \bibinfo {pages} {2080}\BibitemShut {NoStop}%
\bibitem [{\citenamefont {Villars}\ and\ \citenamefont
  {Cenzual}(2012)}]{Villars2012}%
  \BibitemOpen
  \bibinfo {editor} {\bibfnamefont {P.}~\bibnamefont {Villars}}\ and\ \bibinfo
  {editor} {\bibfnamefont {K.}~\bibnamefont {Cenzual}},\ eds.,\ \href {\doibase
  10.1007/978-3-642-22847-6} {\emph {\bibinfo {title} {{Landolt-B{\"{o}}rnstein
  - Group III Condensed Matter, Volume 43A11}}}},\ \bibinfo {series}
  {Landolt-B{\"{o}}rnstein - Group III Condensed Matter}, Vol.\ \bibinfo
  {volume} {43A11}\ (\bibinfo  {publisher} {Springer Berlin Heidelberg},\
  \bibinfo {address} {Berlin, Heidelberg},\ \bibinfo {year} {2012})\BibitemShut
  {NoStop}%
\bibitem [{\citenamefont {Courtois}\ \emph {et~al.}(2008)\citenamefont
  {Courtois}, \citenamefont {Meschke}, \citenamefont {Peltonen},\ and\
  \citenamefont {Pekola}}]{Courtois2008}%
  \BibitemOpen
  \bibfield  {author} {\bibinfo {author} {\bibfnamefont {H.}~\bibnamefont
  {Courtois}}, \bibinfo {author} {\bibfnamefont {M.}~\bibnamefont {Meschke}},
  \bibinfo {author} {\bibfnamefont {J.~T.}\ \bibnamefont {Peltonen}}, \ and\
  \bibinfo {author} {\bibfnamefont {J.~P.}\ \bibnamefont {Pekola}},\ }\href
  {\doibase 10.1103/PhysRevLett.101.067002} {\bibfield  {journal} {\bibinfo
  {journal} {Physical Review Letters}\ }\textbf {\bibinfo {volume} {101}},\
  \bibinfo {pages} {1} (\bibinfo {year} {2008})}\BibitemShut {NoStop}%
\bibitem [{\citenamefont {Tinkham}(2004)}]{Tinkham}%
  \BibitemOpen
  \bibfield  {author} {\bibinfo {author} {\bibfnamefont {M.}~\bibnamefont
  {Tinkham}},\ }\href@noop {} {\emph {\bibinfo {title} {{Introduction to
  Superconductivity Second Edition}}}}\ (\bibinfo  {publisher} {Dover
  Publications, Inc.},\ \bibinfo {address} {Mineola, New York},\ \bibinfo
  {year} {2004})\BibitemShut {NoStop}%
\bibitem [{\citenamefont {Meservey}\ and\ \citenamefont
  {Tedrow}(1971)}]{Meservey1971}%
  \BibitemOpen
  \bibfield  {author} {\bibinfo {author} {\bibfnamefont {R.}~\bibnamefont
  {Meservey}}\ and\ \bibinfo {author} {\bibfnamefont {P.~M.}\ \bibnamefont
  {Tedrow}},\ }\href {\doibase 10.1063/1.1659648} {\bibfield  {journal}
  {\bibinfo  {journal} {Journal of Applied Physics}\ }\textbf {\bibinfo
  {volume} {42}},\ \bibinfo {pages} {51} (\bibinfo {year} {1971})}\BibitemShut
  {NoStop}%
\bibitem [{\citenamefont {Deutscher}\ and\ \citenamefont
  {Rappaport}(1979)}]{Deutscher1979}%
  \BibitemOpen
  \bibfield  {author} {\bibinfo {author} {\bibfnamefont {G.}~\bibnamefont
  {Deutscher}}\ and\ \bibinfo {author} {\bibfnamefont {M.}~\bibnamefont
  {Rappaport}},\ }\href {\doibase 10.1051/jphyslet:019790040010021900}
  {\bibfield  {journal} {\bibinfo  {journal} {Journal de Physique Lettres}\
  }\textbf {\bibinfo {volume} {40}},\ \bibinfo {pages} {219} (\bibinfo {year}
  {1979})}\BibitemShut {NoStop}%
\bibitem [{\citenamefont {Lesueur}\ \emph {et~al.}(1988)\citenamefont
  {Lesueur}, \citenamefont {Dumoulin},\ and\ \citenamefont
  {N{\'{e}}dellec}}]{Lesueur1988}%
  \BibitemOpen
  \bibfield  {author} {\bibinfo {author} {\bibfnamefont {J.}~\bibnamefont
  {Lesueur}}, \bibinfo {author} {\bibfnamefont {L.}~\bibnamefont {Dumoulin}}, \
  and\ \bibinfo {author} {\bibfnamefont {P.}~\bibnamefont {N{\'{e}}dellec}},\
  }\href {\doibase 10.1016/0038-1098(88)90992-1} {\bibfield  {journal}
  {\bibinfo  {journal} {Solid State Communications}\ }\textbf {\bibinfo
  {volume} {66}},\ \bibinfo {pages} {723} (\bibinfo {year} {1988})}\BibitemShut
  {NoStop}%
\bibitem [{\citenamefont {Deutscher}\ \emph {et~al.}(1971)\citenamefont
  {Deutscher}, \citenamefont {Farges}, \citenamefont {Meunier},\ and\
  \citenamefont {Nedellec}}]{Deutscher1971}%
  \BibitemOpen
  \bibfield  {author} {\bibinfo {author} {\bibfnamefont {G.}~\bibnamefont
  {Deutscher}}, \bibinfo {author} {\bibfnamefont {J.}~\bibnamefont {Farges}},
  \bibinfo {author} {\bibfnamefont {F.}~\bibnamefont {Meunier}}, \ and\
  \bibinfo {author} {\bibfnamefont {P.}~\bibnamefont {Nedellec}},\ }\href
  {\doibase 10.1016/0375-9601(71)90374-4} {\bibfield  {journal} {\bibinfo
  {journal} {Physics Letters A}\ }\textbf {\bibinfo {volume} {35}},\ \bibinfo
  {pages} {265} (\bibinfo {year} {1971})}\BibitemShut {NoStop}%
\bibitem [{\citenamefont {Tsuei}\ and\ \citenamefont
  {Johnson}(1974)}]{Tsuei1974}%
  \BibitemOpen
  \bibfield  {author} {\bibinfo {author} {\bibfnamefont {C.~C.}\ \bibnamefont
  {Tsuei}}\ and\ \bibinfo {author} {\bibfnamefont {W.~L.}\ \bibnamefont
  {Johnson}},\ }\href {\doibase 10.1103/PhysRevB.9.4742} {\bibfield  {journal}
  {\bibinfo  {journal} {Physical Review B}\ }\textbf {\bibinfo {volume} {9}},\
  \bibinfo {pages} {4742} (\bibinfo {year} {1974})}\BibitemShut {NoStop}%
\bibitem [{\citenamefont {Kuan}\ \emph {et~al.}(1982)\citenamefont {Kuan},
  \citenamefont {Chen}, \citenamefont {Yi}, \citenamefont {Wang}, \citenamefont
  {Wu},\ and\ \citenamefont {Garoche}}]{Kuan1982}%
  \BibitemOpen
  \bibfield  {author} {\bibinfo {author} {\bibfnamefont {W.}~\bibnamefont
  {Kuan}}, \bibinfo {author} {\bibfnamefont {S.}~\bibnamefont {Chen}}, \bibinfo
  {author} {\bibfnamefont {S.}~\bibnamefont {Yi}}, \bibinfo {author}
  {\bibfnamefont {Z.}~\bibnamefont {Wang}}, \bibinfo {author} {\bibfnamefont
  {C.}~\bibnamefont {Wu}}, \ and\ \bibinfo {author} {\bibfnamefont
  {P.}~\bibnamefont {Garoche}},\ }\href {\doibase 10.1007/BF00655053}
  {\bibfield  {journal} {\bibinfo  {journal} {Journal of Low Temperature
  Physics}\ }\textbf {\bibinfo {volume} {46}},\ \bibinfo {pages} {237}
  (\bibinfo {year} {1982})}\BibitemShut {NoStop}%
\bibitem [{\citenamefont {Chevrier}\ \emph {et~al.}(1987)\citenamefont
  {Chevrier}, \citenamefont {Pavuna},\ and\ \citenamefont
  {Cyrot-Lackmann}}]{Chevrier1987}%
  \BibitemOpen
  \bibfield  {author} {\bibinfo {author} {\bibfnamefont {J.}~\bibnamefont
  {Chevrier}}, \bibinfo {author} {\bibfnamefont {D.}~\bibnamefont {Pavuna}}, \
  and\ \bibinfo {author} {\bibfnamefont {F.}~\bibnamefont {Cyrot-Lackmann}},\
  }\href {\doibase 10.1103/PhysRevB.36.9115} {\bibfield  {journal} {\bibinfo
  {journal} {Physical Review B}\ }\textbf {\bibinfo {volume} {36}},\ \bibinfo
  {pages} {9115} (\bibinfo {year} {1987})}\BibitemShut {NoStop}%
\bibitem [{\citenamefont {Xi}\ \emph {et~al.}(1987)\citenamefont {Xi},
  \citenamefont {Ran}, \citenamefont {Liu},\ and\ \citenamefont
  {Guan}}]{Xi1987}%
  \BibitemOpen
  \bibfield  {author} {\bibinfo {author} {\bibfnamefont {X.-X.}\ \bibnamefont
  {Xi}}, \bibinfo {author} {\bibfnamefont {Q.-Z.}\ \bibnamefont {Ran}},
  \bibinfo {author} {\bibfnamefont {J.-R.}\ \bibnamefont {Liu}}, \ and\
  \bibinfo {author} {\bibfnamefont {W.-Y.}\ \bibnamefont {Guan}},\ }\href
  {\doibase 10.1016/0038-1098(87)90479-0} {\bibfield  {journal} {\bibinfo
  {journal} {Solid State Communications}\ }\textbf {\bibinfo {volume} {61}},\
  \bibinfo {pages} {791} (\bibinfo {year} {1987})}\BibitemShut {NoStop}%
\bibitem [{\citenamefont {Kim}\ \emph {et~al.}(2013)\citenamefont {Kim},
  \citenamefont {Ahn}, \citenamefont {Kim}, \citenamefont {Choi}, \citenamefont
  {Bae}, \citenamefont {Kang}, \citenamefont {Lim}, \citenamefont
  {L{\'{o}}pez},\ and\ \citenamefont {Kim}}]{Kim2013}%
  \BibitemOpen
  \bibfield  {author} {\bibinfo {author} {\bibfnamefont {B.-K.}\ \bibnamefont
  {Kim}}, \bibinfo {author} {\bibfnamefont {Y.-H.}\ \bibnamefont {Ahn}},
  \bibinfo {author} {\bibfnamefont {J.-J.}\ \bibnamefont {Kim}}, \bibinfo
  {author} {\bibfnamefont {M.-S.}\ \bibnamefont {Choi}}, \bibinfo {author}
  {\bibfnamefont {M.-H.}\ \bibnamefont {Bae}}, \bibinfo {author} {\bibfnamefont
  {K.}~\bibnamefont {Kang}}, \bibinfo {author} {\bibfnamefont {J.~S.}\
  \bibnamefont {Lim}}, \bibinfo {author} {\bibfnamefont {R.}~\bibnamefont
  {L{\'{o}}pez}}, \ and\ \bibinfo {author} {\bibfnamefont {N.}~\bibnamefont
  {Kim}},\ }\href {\doibase 10.1103/PhysRevLett.110.076803} {\bibfield
  {journal} {\bibinfo  {journal} {Physical Review Letters}\ }\textbf {\bibinfo
  {volume} {110}},\ \bibinfo {pages} {076803} (\bibinfo {year}
  {2013})}\BibitemShut {NoStop}%
\bibitem [{\citenamefont {Pillet}\ \emph {et~al.}(2010)\citenamefont {Pillet},
  \citenamefont {Quay}, \citenamefont {Morfin}, \citenamefont {Bena},
  \citenamefont {Yeyati},\ and\ \citenamefont {Joyez}}]{Pillet2010}%
  \BibitemOpen
  \bibfield  {author} {\bibinfo {author} {\bibfnamefont {J.-D.}\ \bibnamefont
  {Pillet}}, \bibinfo {author} {\bibfnamefont {C.~H.~L.}\ \bibnamefont {Quay}},
  \bibinfo {author} {\bibfnamefont {P.}~\bibnamefont {Morfin}}, \bibinfo
  {author} {\bibfnamefont {C.}~\bibnamefont {Bena}}, \bibinfo {author}
  {\bibfnamefont {A.~L.}\ \bibnamefont {Yeyati}}, \ and\ \bibinfo {author}
  {\bibfnamefont {P.}~\bibnamefont {Joyez}},\ }\href {\doibase
  10.1038/nphys1811} {\bibfield  {journal} {\bibinfo  {journal} {Nature
  Physics}\ }\textbf {\bibinfo {volume} {6}},\ \bibinfo {pages} {965} (\bibinfo
  {year} {2010})}\BibitemShut {NoStop}%
\bibitem [{\citenamefont {Gramich}\ \emph {et~al.}(2017)\citenamefont
  {Gramich}, \citenamefont {Baumgartner},\ and\ \citenamefont
  {Sch{\"{o}}nenberger}}]{Gramich2017a}%
  \BibitemOpen
  \bibfield  {author} {\bibinfo {author} {\bibfnamefont {J.}~\bibnamefont
  {Gramich}}, \bibinfo {author} {\bibfnamefont {A.}~\bibnamefont
  {Baumgartner}}, \ and\ \bibinfo {author} {\bibfnamefont {C.}~\bibnamefont
  {Sch{\"{o}}nenberger}},\ }\href {\doibase 10.1103/PhysRevB.96.195418}
  {\bibfield  {journal} {\bibinfo  {journal} {Physical Review B}\ }\textbf
  {\bibinfo {volume} {96}},\ \bibinfo {pages} {1} (\bibinfo {year}
  {2017})}\BibitemShut {NoStop}%
\bibitem [{\citenamefont {Beenakker}(1992)}]{Beenakker1992a}%
  \BibitemOpen
  \bibfield  {author} {\bibinfo {author} {\bibfnamefont {C.~W.~J.}\
  \bibnamefont {Beenakker}},\ }\href {\doibase 10.1103/PhysRevB.46.12841}
  {\bibfield  {journal} {\bibinfo  {journal} {Physical Review B}\ }\textbf
  {\bibinfo {volume} {46}},\ \bibinfo {pages} {12841} (\bibinfo {year}
  {1992})}\BibitemShut {NoStop}%
\bibitem [{\citenamefont {Gross}\ and\ \citenamefont {Marx}(2005)}]{Gross2005}%
  \BibitemOpen
  \bibfield  {author} {\bibinfo {author} {\bibfnamefont {R.}~\bibnamefont
  {Gross}}\ and\ \bibinfo {author} {\bibfnamefont {A.}~\bibnamefont {Marx}},\
  }\href@noop {} {\emph {\bibinfo {title} {{Applied superconductivity,}}}},\
  Vol.~\bibinfo {volume} {1}\ (\bibinfo {year} {2005})\ p.\ \bibinfo {pages}
  {480}\BibitemShut {NoStop}%
\bibitem [{\citenamefont {Maier}\ \emph
  {et~al.}(2014{\natexlab{b}})\citenamefont {Maier}, \citenamefont {Meng},\
  and\ \citenamefont {Loss}}]{Maier2014a}%
  \BibitemOpen
  \bibfield  {author} {\bibinfo {author} {\bibfnamefont {F.}~\bibnamefont
  {Maier}}, \bibinfo {author} {\bibfnamefont {T.}~\bibnamefont {Meng}}, \ and\
  \bibinfo {author} {\bibfnamefont {D.}~\bibnamefont {Loss}},\ }\href {\doibase
  10.1103/PhysRevB.90.155437} {\bibfield  {journal} {\bibinfo  {journal}
  {Physical Review B}\ }\textbf {\bibinfo {volume} {90}},\ \bibinfo {pages}
  {155437} (\bibinfo {year} {2014}{\natexlab{b}})}\BibitemShut {NoStop}%
\bibitem [{\citenamefont {Stanescu}\ and\ \citenamefont
  {Tewari}(2013)}]{Stanescu2013a}%
  \BibitemOpen
  \bibfield  {author} {\bibinfo {author} {\bibfnamefont {T.~D.}\ \bibnamefont
  {Stanescu}}\ and\ \bibinfo {author} {\bibfnamefont {S.}~\bibnamefont
  {Tewari}},\ }\href {\doibase 10.1088/0953-8984/25/23/233201} {\bibfield
  {journal} {\bibinfo  {journal} {Journal of Physics: Condensed Matter}\
  }\textbf {\bibinfo {volume} {25}},\ \bibinfo {pages} {233201} (\bibinfo
  {year} {2013})}\BibitemShut {NoStop}%
\end{thebibliography}
%

\pagebreak
\pagebreak
\widetext
\begin{center}
\textbf{\large Supporting Information: \\ Hard superconducting gap and diffusion-induced superconductors in Ge-Si nanowires}
\end{center}
\setcounter{equation}{0}
\setcounter{figure}{0}
\setcounter{table}{0}
\setcounter{section}{0}
\makeatletter
\renewcommand{\thesection}{S\Roman{section}}
\renewcommand{\theequation}{S\arabic{equation}}
\renewcommand{\thefigure}{S\arabic{figure}}
\renewcommand{\thetable}{S\arabic{table}}
\renewcommand{\bibnumfmt}[1]{[S#1]}
\renewcommand{\citenumfont}[1]{S#1}

\section{Fabrication}
\label{fabrication}
Ge-Si core-shell nanowires are deposited on a \emph{p}$^{++}$ doped Si substrate covered with $100$~nm \sio and contacted after AFM imaging. Source and drain contacts are defined using electron-beam lithograph and after developing, a 3 second buffered hydrofluoric acid ($12.5$~\%) dip is performed to remove native \sio from the Si shell of the nanowire. The contacts are metallized using electron-beam evaporation of Al, resulting in a $150$~nm nanowire channel. As a last step, devices are annealed for $10$~minutes on a hotplate in ambient at $180^\circ$C during which Al diffuses into the wire. As a result, a drop in room temperature resistance from several M$\Omega$ to several k$\Omega$ was observed for $\sim80$~\% of devices.
\paragraph*{}
In total, 7 out of 15 devices tested at low temperature showed a supercurrent with $4$ devices showing a superconducting phase with a $\tc$ between $600$ and $900$ mK, comparable to X$1$. Three devices showed a metallized nanowire with comparable $\ic$ to device B, we have no information on their respective $\tc$ or $\bc$.

\section{Extraction of $\tc$-$\bcs$ curves}
\label{curveextract}
Here we explain how the $\tc$-$\bcs$ curves of Fig.~3b in the main text were generated. We define $\bcsal$ ($\bcsx$) as the field where Al (X$1$) no longer induces a supercurrent, \ie, we no longer observe a $\ic$. In Fig.~3a we cannot directly observe $\bcsal$ for $0<T<800$~mK where the `peak' of Al and the `tail' of X$1$ overlap (for $230<\bs<300$ mT) and we therefore use the following method: (1) For all temperatures we take each individual $\ic$-$\bs$ curve, \ie, horizontal linecuts in Fig.3a, (2) we select only $\ic$ where $\bs<\bcsal$\ (3) we fit $\ic$ to an empirical polynomial of the form $\ic(\bs)=a\bs^4+b\bs^2+c$ with $a$ and $b$ only allowed negative while $c$ is always positive, (4) we find $\bcsal$ for each temperature as the roots of $\ic(\bs)$ (\ie, zero crossings). For X$1$ we use the same method except in (2) we select only $\ic$ for $\bs>\bcsal$. 

\clearpage
\section{Additional figures and tables}
\begin{figure*}[!htb]
  \includegraphics{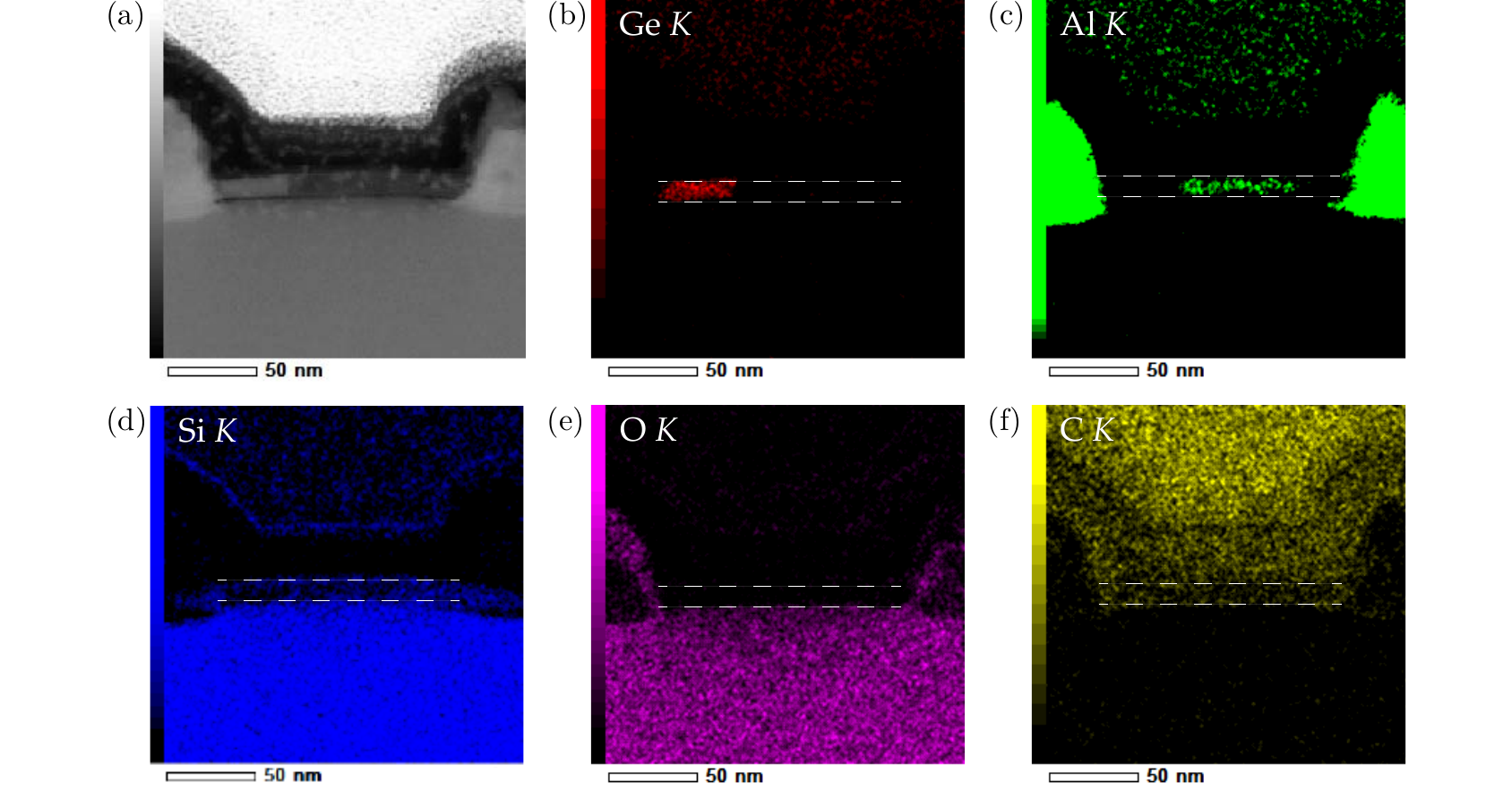}
  \caption{\textbf{EDX maps of various elements of device A:} a) TEM image of device A, same as Fig.1b in the main text. EDX map for Ge (b), Al (c), Si (d), O (e) and C (f). White dashed lines indicate the approximate position of the nanowire. (a), (b), (c) and (d) were used to construct Fig.1c in the main text.}
  \label{SI-4}
\end{figure*}

\begin{figure*}
  \includegraphics{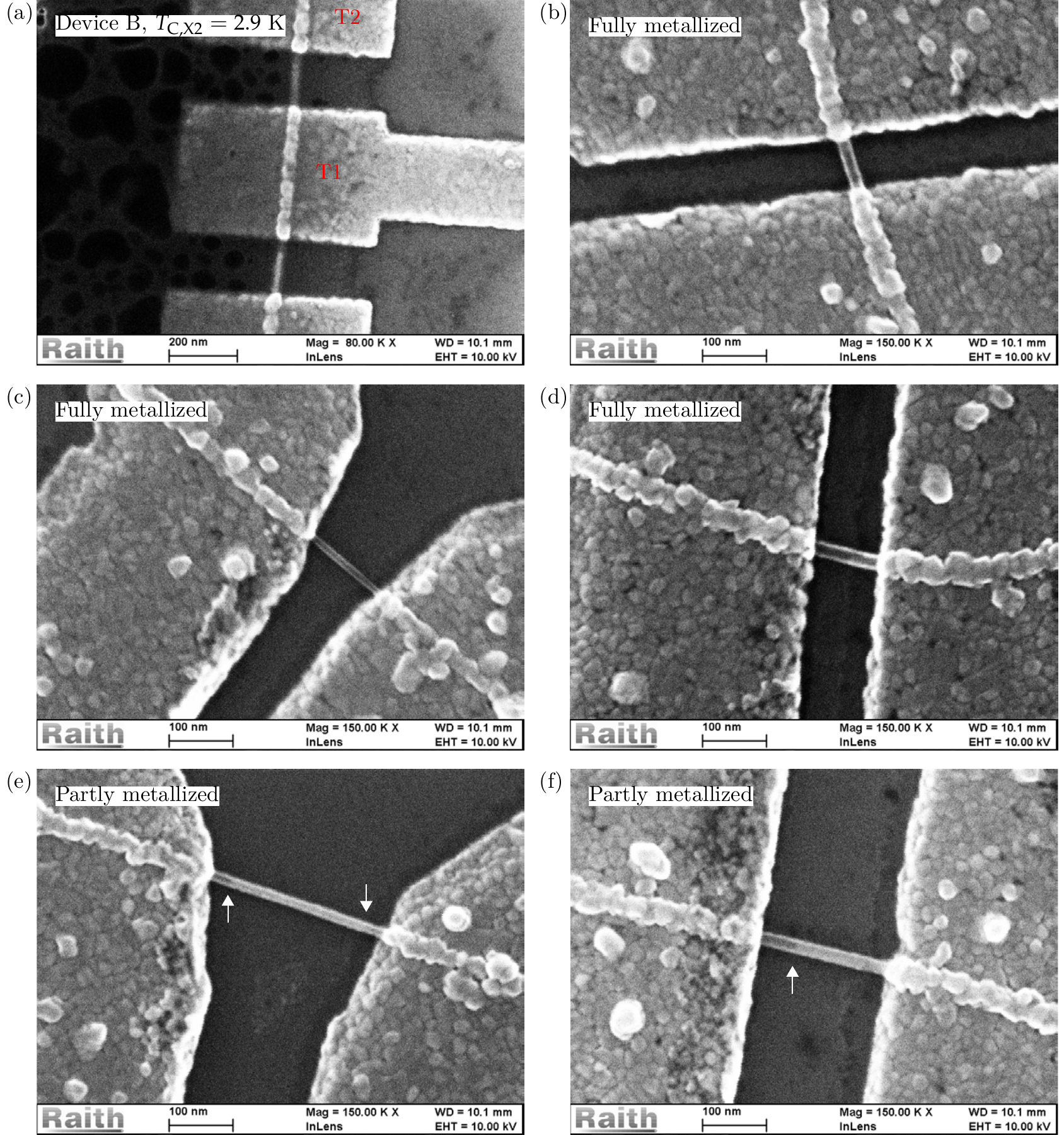}
  \caption{\textbf{SEM images of devices:} All devices have undergone the same annealing process. a) SEM image of device B. Carbon contamination prevents a conclusive analysis of the Al-Ge inter-diffusion process. We suspect that the metallized nanowire segment is located between T1 and T2 based on the low contrast of the nanowire segment. b) Another metallized device exhibiting the same darkened color in the nanowire segment as in a), no additional superconducting phase was found in this device. c), d) More devices where the complete channel shows a low contrast for which we suspect they are fully metallized. e), f) Devices with a region of lower contrast close to the Al contacts, suspected due to partial metalization with the white arrows denoting the Al-Ge inter-diffusion front. Devices c) - f) were not measured. Other devices that were measured could not be imaged afterwards.}
  \label{SI-5}
\end{figure*}

\begin{figure*}
  \includegraphics{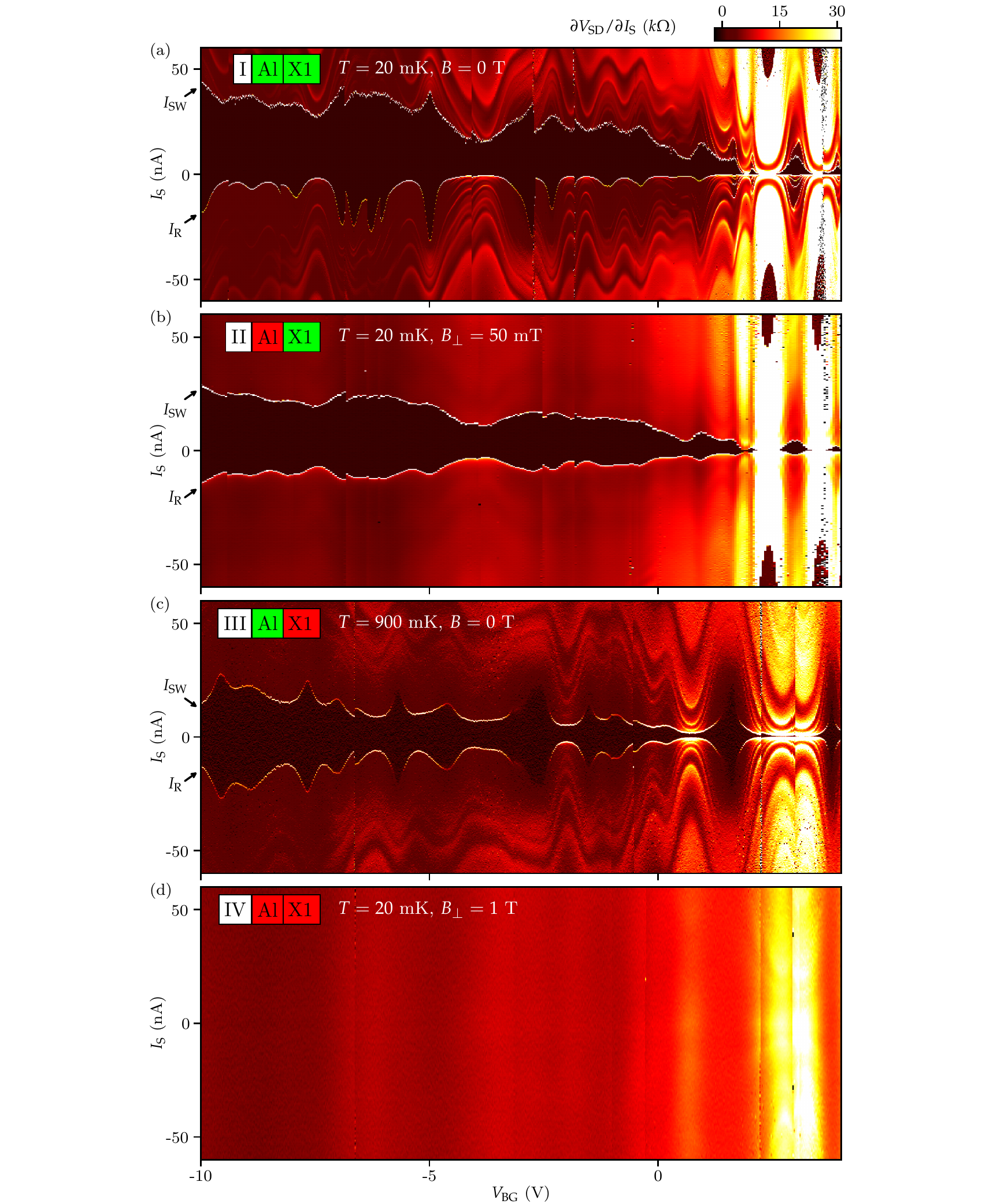}
  \caption{\textbf{Backgate dependence for all four configurations:} All figures show $\dvdi$ vs $\is$ and $\vbg$ with $\is$ measured from negative towards positive bias. a) Configuration~\textbf{I} taken at $T=20$~mK and $B=0$~T. b) Configuration~\textbf{II} taken at $T=20$~mK and $\bpe=50$~mT. c) Configuration~\textbf{III} taken at $T=900$~mK and $B=0$~T. d) Configuration~\textbf{IV} taken at $T=20$~mK and $\bpe=1$~T. In all figures, $\is$ was swept from negative to positive bias during measurement. a), b) and d) were taken during the same cooldown, all figures share the same color scale displayed in a). In a), b) and c), $\ic$ and $\ir$ are denoted.}
  \label{SI-10}
\end{figure*}

\begin{figure*}
  \includegraphics{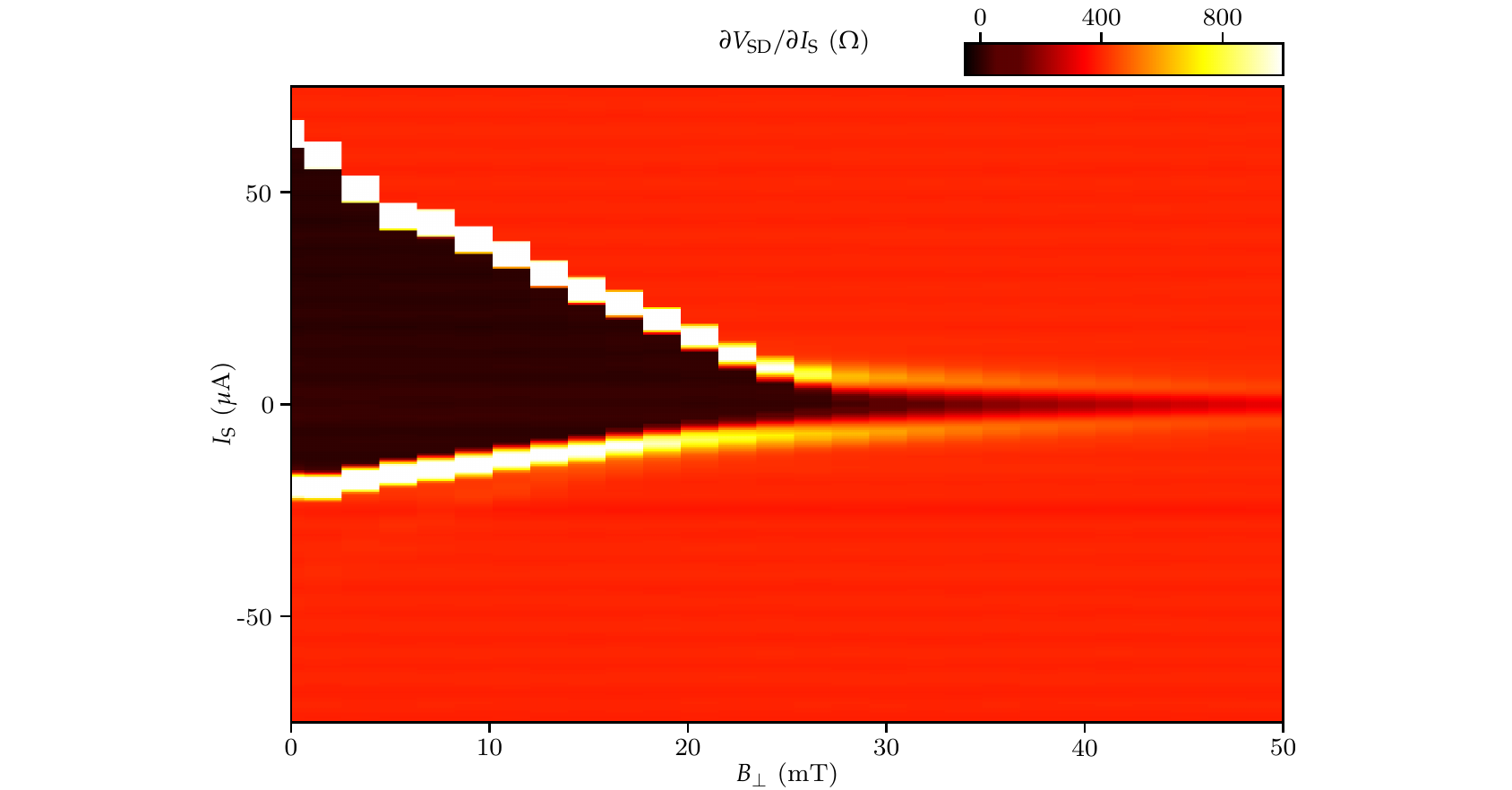}
  \caption{\textbf{Al lead in an out-of-plane field:} $\dvdi$ vs $\is$ and $\bpe$. The Al lead has a thickness of $\sim50$~nm and $\ic$ is determined by its smallest width of $\sim500$~nm running over a length of several $\mu$m. The black region denotes a supercurrent. The lead is measured in a 2-probe configuration and a series resistance of two times the line resistance is subtracted. Measurements are taken in the positive bias direction where the asymmetry in bias is attributed to local Joule heating.}
  \label{SI-2-5}
\end{figure*}

\begin{figure*}[ht!]
  \includegraphics{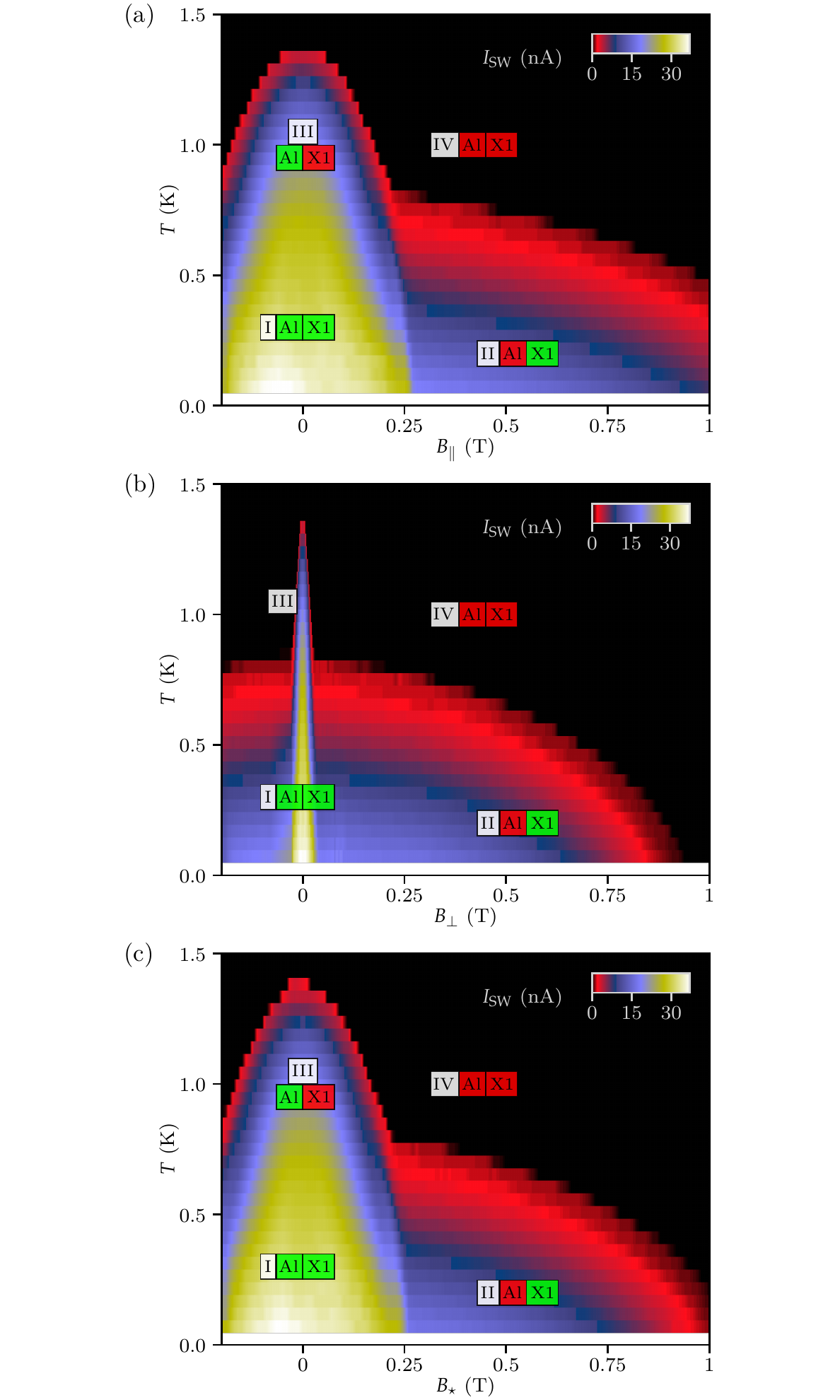}
  \caption{\textbf{$\ic$ vs $T$ and $B$ of Device A:} Datasets were used to generate Fig.3b in the main text as explained in section~SII. a) $\ic$ vs $\bpa$ and $T$, b) $\ic$ vs $\bpe$ and $T$ and c) $\ic$ vs $\bs$ vs T (same as Fig.3a in the main text). Configuration I-IV are denoted in all figures in which green indicates superconductivity while red indicates the normal state.}
  \label{SI-1}
\end{figure*}

\begin{table*}[p]
\centering
\caption{Configurations of device A with two superconducting phases and the corresponding conditions for the magnetic field $\mathbf{B}$ and temperature $\mathbf{T}$, SC refers to superconductivity. The last column refers to the plots in Fig.2d. For generality the field direction is removed from the subscripts.}
\label{hgconfs}
	\begin{ruledtabular}
\begin{tabular}{c>{\columncolor[RGB]{3, 191, 61}}c>{\columncolor[RGB]{255, 5, 0}}ccc}
\rowcolor{white!50}
\textbf{Configuration} & \textbf{Superconducting}    & \textbf{Normal state} & \textbf{Conditions} & \textbf{$\mathbf{\ic}$ and symbol} \\ \hline
\textbf{I}    & Al,X$1$ &        & $\mathbf{B}<\bcal,\bcx$      & $\ic\approx36$~nA 		\\
\rowcolor{white!50} & &        & \&  $\mathbf{T}<\tcx,\tcal$	  & $\color{NavyBlue} \medblackcircle$         \\ \hline
\textbf{II}   & X$1$    & Al     & $\bcal<\mathbf{B}<\bcx$      & $\ic\approx17$~nA      \\
\rowcolor{white!50} & &        & \&  $\mathbf{T}<\tcx,\tcal$   & $\color{DarkOrchid} \medblacktriangleup$       \\ \hline
\textbf{III}  & Al    & X$1$     & $\mathbf{B}<\bcal,\bcx$      & $\ic\approx24$~nA      \\
\rowcolor{white!50} & &        & \&  $\tcx<\mathbf{T}<\tcal$   & $\color{Brown} \medblacksquare$        \\ \hline
\textbf{IV}   &       & Al,X$1$  & $\mathbf{B}>\bcx$            & $\ic=0$          \\
\rowcolor{white!50} & &        &or  $\mathbf{T}>\tcal$         & $\color{Gray} \medblacktriangledown $        \\
\rowcolor{white!50} & &        &or  $\bcal<\mathbf{B}<\bcx$    &              \\
\rowcolor{white!50} & &        &\&  $\tcx<\mathbf{T}<\tcal$    &     			\\       
\end{tabular}
\end{ruledtabular}
\end{table*}

\begin{figure*}
  \includegraphics{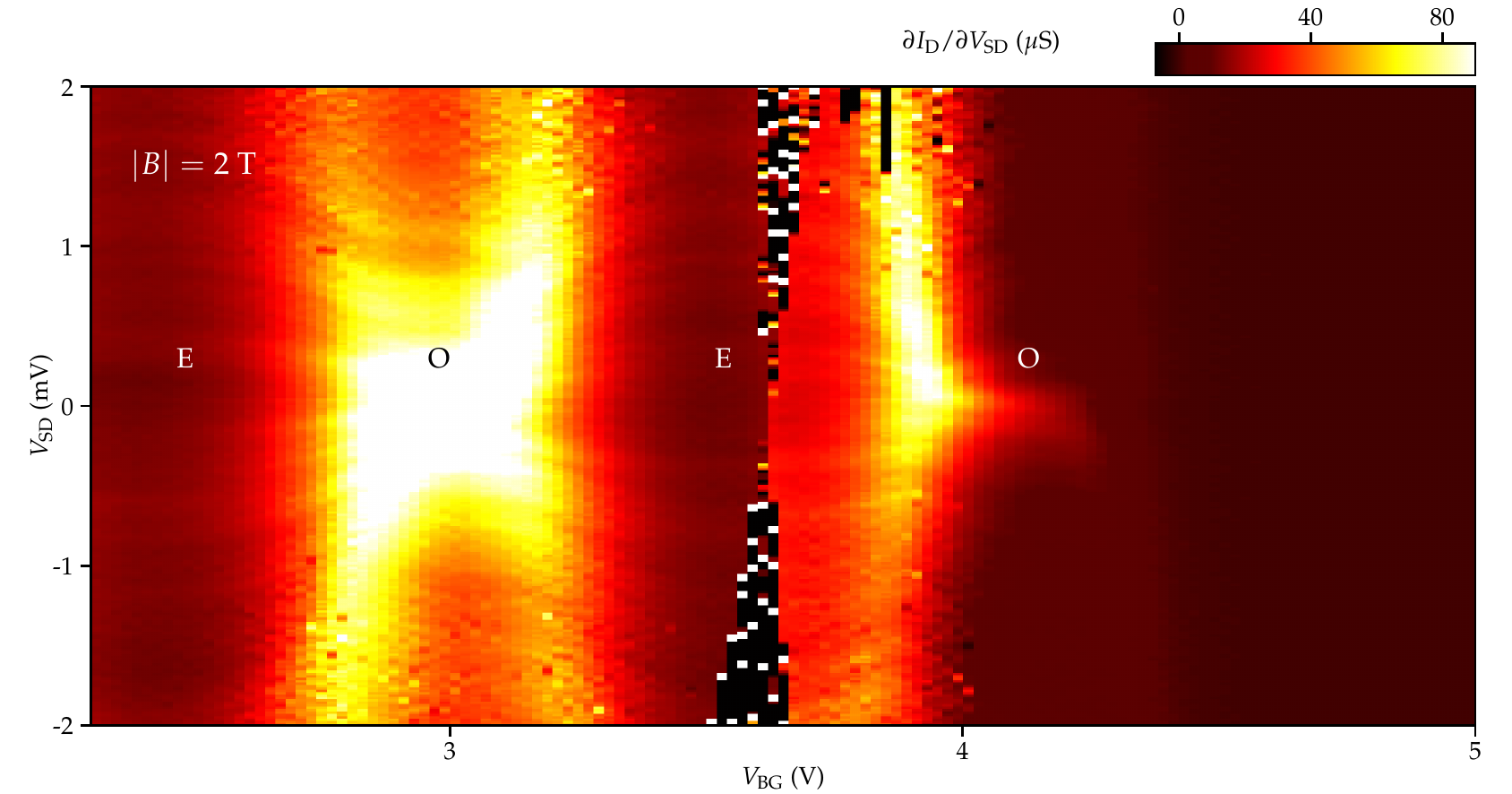}
  \caption{\textbf{Kondo effect at finite magnetic field:} $\didvb$ vs $\vsd$ and $\vbg$ for $|B|=2$~T. Odd (O) and even (E) hole occupation of the nanowire quantum dot is indicated. The Kondo effect is visible for odd occupation as increased conductance around $\vsd=0$.}
  \label{SI-2}
\end{figure*}

\begin{figure*}
  \includegraphics{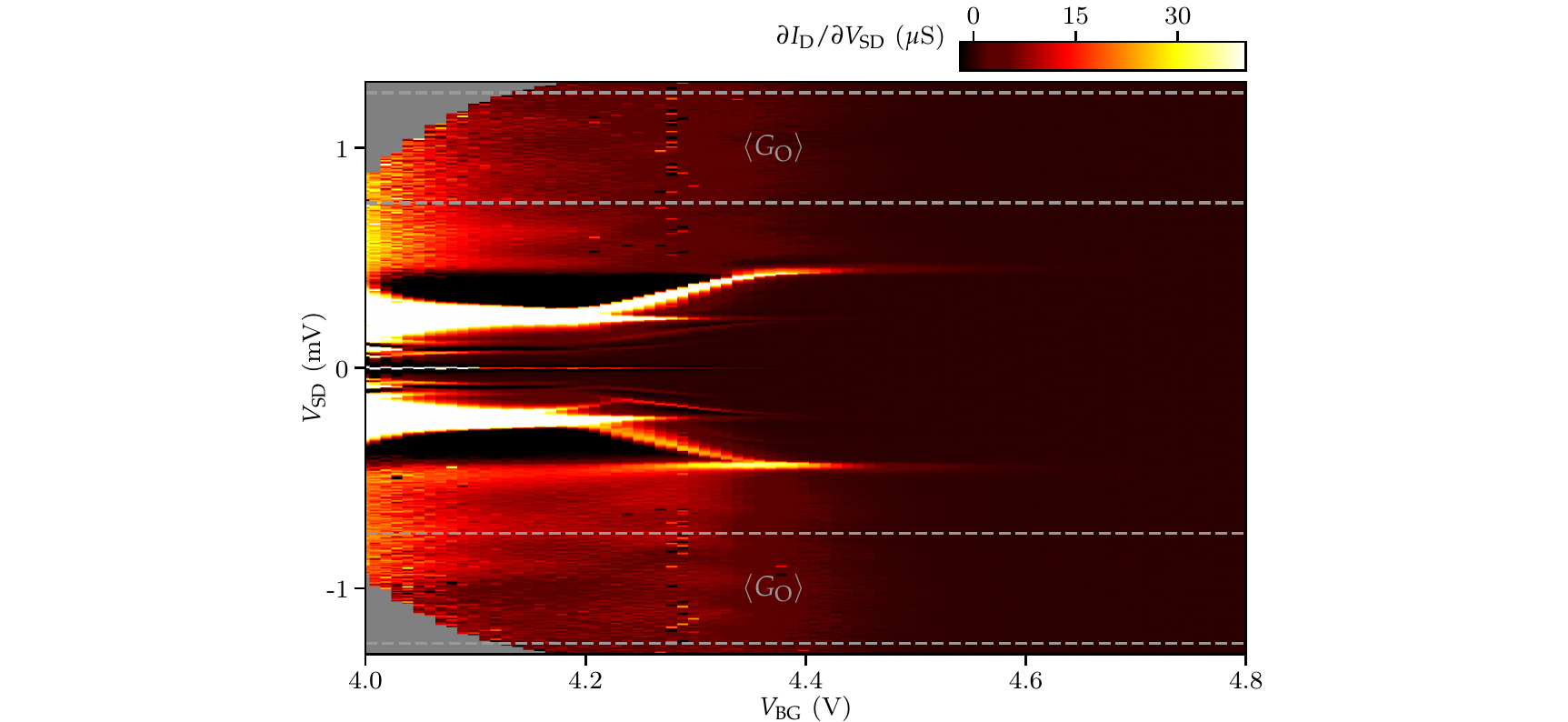}
  \caption{\textcolor{black}{\textbf{Determination of $\langle G_\mathrm{O}\rangle$:} $\langle G_\mathrm{O}\rangle$ is determined from this dataset measured at higher bias compared to the dataset shown in Fig.4a in the main text. $\langle G_\mathrm{O}\rangle$ is determined as the averaged conductance over a range $\vsd$ delimited by the dashed grey lines. Conductance values for positive and negative bias are used. A correction to their corresponding values of $\vbg$ is applied to compensate for the gate lever arm $\alpha\approx0.18$ (see Ref.[34] in the main text).}}
  \label{SI-GO}
\end{figure*}

\begin{figure*}
  \includegraphics{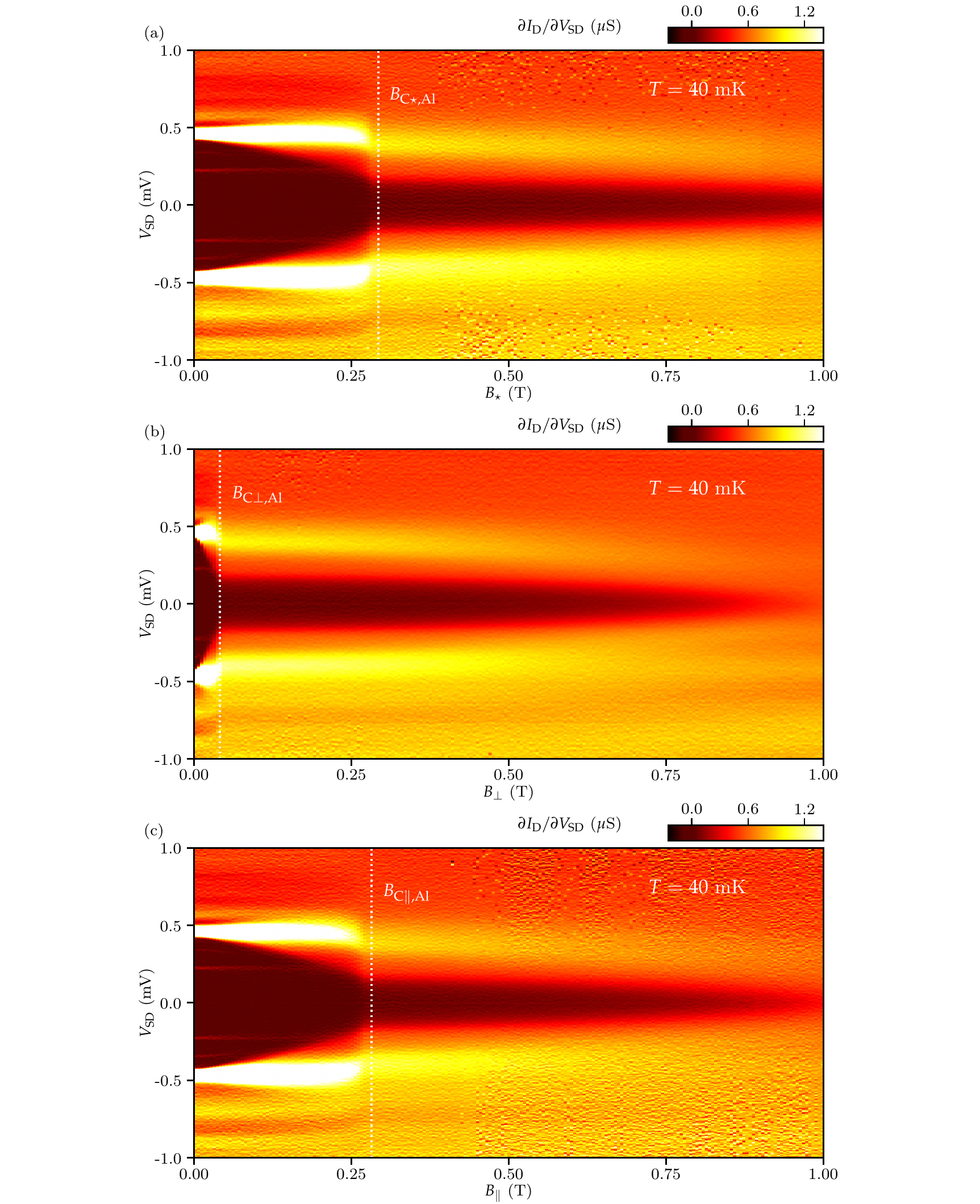}
  \caption{\textbf{Superconducting gap versus $B$:} Datasets used to generate Fig.4d and Fig.4e in the main text, taken at $\vbg=4.45$ (blue dotted line in Fig.a. a) $\didvb$ vs $\vsd$ and $B_\star$, the white dotted line indicates the critical field of Al $\bcsal$. b) $\didvb$ vs $\vsd$ and $B_\perp$, the white dotted line indicates $\bcpeal$. c) $\didvb$ vs $\vsd$ and $B_\parallel$, the white dotted line indicates $\bcpaal$. Faint zero-bias peaks attributed to the Kondo effect appear only above $\bcsal$.}
  \label{SI-6}
\end{figure*}

\begin{figure*}
  \includegraphics{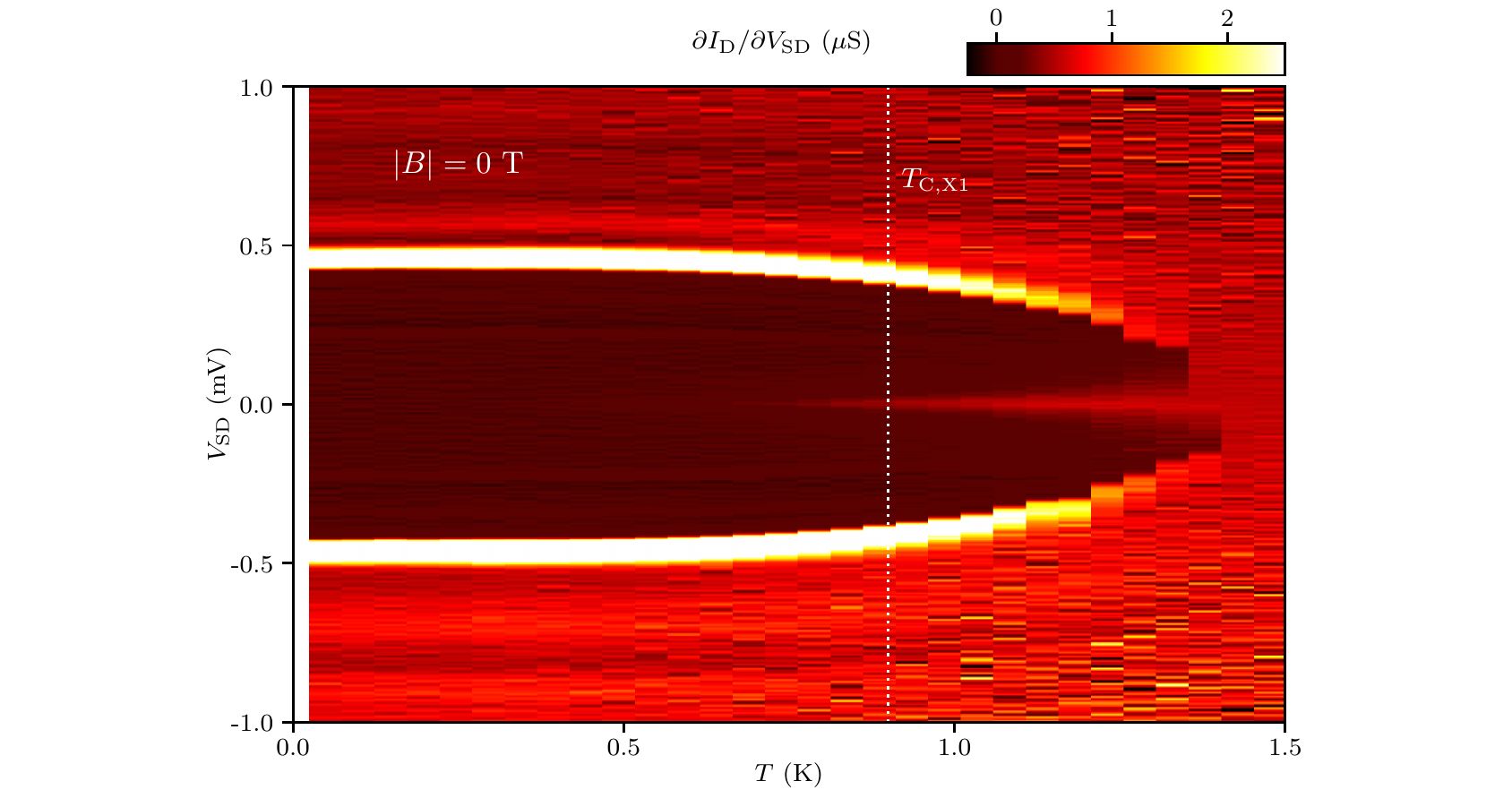}
  \caption{\textbf{Superconducting gap versus $T$:} $\didvb$ vs $\vsd$ and $T$ taken at $\vbg=4.45$ (blue dotted line in Fig.4a). The white dotted line indicates the critical temperature of X$1$ $\tcx$. No abrubt change in conductance is observed above $\tcx$. }
  \label{SI-3}
\end{figure*}

\end{document}